\documentclass[journal]{IEEEtran}
\usepackage{amsfonts}
\usepackage[cmex10]{amsmath}
\usepackage{cite}
\usepackage{graphicx}
\newtheorem{theorem}{Theorem}[section]
\newtheorem{corollary}[theorem]{Corollary}
\newtheorem{lemma}[theorem]{Lemma}
\newtheorem{notation}[theorem]{Notation}
\newtheorem{example}[theorem]{Example}

\ifCLASSINFOpdf
\else
\fi
\begin{document}
\bstctlcite{IEEEexample:BSTcontrol}
%
\title{Self-dual binary $[8m, 4m]$-codes constructed by
left ideals of the dihedral group algebra $\mathbb{F}_2[D_{8m}]$}
%
%
%

\author{Yuan Cao,
        Yonglin Cao, Fang-Wei Fu
        and~Jian Gao
\thanks{Yuan Cao is with School of Mathematics and Statistics, Shandong University of Technology, Zibo, Shandong 255091, China. e-mail: yuancao@sdut.edu.cn.}
\thanks{Yonglin Cao (Corresponding author) is with School of Mathematics and Statistics, Shandong University of Technology, Zibo, Shandong 255091, China. e-mail: ylcao@sdut.edu.cn.}
\thanks{Fang-Wei Fu is with Chern Institute of Mathematics and LPMC, Nankai University, Tianjin 300071, China. e-mail: fwfu@nankai.edu.cn.}
\thanks{Jian Gao is with School of Mathematics and Statistics, Shandong University of Technology, Zibo, Shandong 255091, China. e-mail: dezhougaojian@163.com.}
\thanks{Copyright (c) 2014 IEEE. Personal use of this material is permitted.  However, permission to use this material for any other purposes must be obtained from the IEEE by sending a request to pubs-permissions@ieee.org.}
\thanks{Manuscript received  , 2018; revised , .}}

%
%

\markboth{IEEE Transactions on Information Theory,~Vol.~ , No.~ ,  ~ }%
{Shell \MakeLowercase{\textit{et al.}}: Y. Cao, Self-dual binary $[8m, 4m]$-codes}
%



\maketitle

\begin{abstract}
Let $m$ be an arbitrary positive integer and $D_{8m}$ be a dihedral group of order $8m$, i.e., $D_{8m}=\langle x,y\mid x^{4m}=1, y^2=1, yxy=x^{-1}\rangle$. Left ideals of the dihedral group algebra $\mathbb{F}_2[D_{8m}]$ are called
binary left dihedral codes of length $8m$, and abbreviated as binary left $D_{8m}$-codes.
In this paper, we give an explicit representation and enumeration for
all distinct self-dual binary left $D_{8m}$-codes. These codes make up an important class of self-dual binary $[8m,4m]$-codes
such that the dihedral group $D_{8m}$ is necessary a subgroup of the automorphism group of each code. In particular,
we provide recursive algorithms to solve
congruence equations over finite chain rings for constructing all distinct self-dual binary left $D_{8m}$-codes and obtain a Mass formula to count the number of all these self-dual codes. As a preliminary application,
we obtain the extremal self-dual binary $[48,24,12]$-code and an extremal self-dual binary $[56,28,12]$-code
from self-dual binary left $D_{48}$-codes and left $D_{56}$-codes respectively.
\end{abstract}


%
\IEEEpeerreviewmaketitle

\section{Introduction}
\label{intro}
\IEEEPARstart{T}{he}
class of self-dual codes is an interesting topic in coding theory due to
their connections to other fields of mathematics such as Lattices, Cryptography, Invariant Theory, Block designs, etc.
A common theme for the construction of self-dual codes is the use of a computer search. In order to make this search feasible, special construction methods have been used to reduce the search field.
In recent years, one of the important construction methods is to use left ideals in
a finite group algebra over finite fields and finite rings.

\par
  For example, McLoughlin \cite{Mcloughlin2012A} provided a construction of the self-dual, doubly-even and extremal $[48,24,12]$
binary linear block code using a
zero divisor in the dihedral group algebra $\mathbb{F}_2[D_{48}]$.
Dougherty et al. \cite{Dougherty2016Constructions} and \cite{Dougherty2018Group} gave constructions of self-dual and formally self-dual codes from group rings $R[G]$  where the ring $R$ is a finite commutative Frobenius ring. They shown that several of the standard constructions of self-dual codes are found within this general framework.  Additionally, they showed precisely which groups can be used to construct the extremal Type II codes of length 24 and 48.

\par
   A linear code is said to be \textit{self-dual} if $C=C^{\bot}$.
Binary self-dual codes are called Type II if the weights of all codewords are
multiple of $4$ and Type I otherwise. Type II codes are said to have
weights that are doubly-even as well. It is well-known that the upper bound for minimum distance
$d$ of a binary self-dual code of length $n$ is
$$d\leq\left\{\begin{array}{ll}4\lfloor \frac{n}{24}\rfloor+6, & {\rm if} \ n\equiv 22 \ ({\rm mod} \ 24); \cr
 4\lfloor \frac{n}{24}\rfloor+4, & {\rm otherwise}. \end{array}\right.$$
A self-dual binary code is called \textit{extremal} if it meets the bound.

\par
 Let $\mathbb{F}_q$ be a finite field of $q$ elements and $G$ be an arbitrary finite group.
The group algebra $\mathbb{F}_q[G]$ is an $\mathbb{F}_q$-algebra with basis $G$. Addition, multiplication with
scalars $c\in \mathbb{F}_q$ and multiplication are defined by:
$$\sum_{g\in G}a_g g+\sum_{g\in G}b_g g=\sum_{g\in G}(a_g+b_g) g, \
c(\sum_{g\in G}a_g g)=\sum_{g\in G}ca_g g,$$
$$(\sum_{g\in G}a_g g)(\sum_{g\in G}b_g g)=\sum_{g\in G}(\sum_{uv=g}a_ub_v)g,$$
for any $a_g,b_g\in \mathbb{F}_q$ and $g\in G$. Then $\mathbb{F}_q[G]$ is a noncommutative ring with identity $1=1_{\mathbb{F}_q}1_{G}$ where
$1_{\mathbb{F}_q}$ and $1_{G}$ is the identity elements of $\mathbb{F}_q$ and $G$ respectively. It is known that
$\mathbb{F}_q[G]$ is semisimple if and only if ${\rm gcd}(q,|G|)=1$.

\par
   In this paper, let
\begin{eqnarray*}
D_{2n}&=&\langle x,y\mid x^n=1, y^2=1, yxy=x^{-1}\rangle\\
 &=&\{x^iy^j\mid 0\leq i\leq n-1, j=0,1\}
\end{eqnarray*}
 be a dihedral group of order $2n$. For any $a=(a_{0,0}, a_{1,0},\ldots$, $a_{n-1,0},a_{0,1}, a_{1,1}, \ldots, a_{n-1,1})\in \mathbb{F}_q^{2n}$, we define
$$\Psi(a)=\sum_{i=0}^{n-1}a_{i,0}x^i+\sum_{i=0}^{n-1}a_{i,1}x^iy.$$
Then $\Psi$ is an isomorphism of $\mathbb{F}_q$-linear spaces from $\mathbb{F}_q^{2n}$ onto
$\mathbb{F}_q[D_{2n}]$. As a natural generalization of Dutra et al. \cite{Dutra2009Semisimple}, a nonempty subset $\mathcal{C}$ of  $\mathbb{F}_q^{2n}$
is called a \textit{left dihedral code} (or \textit{left $D_{2n}$-code} for more clear) over $\mathbb{F}_q$ if $\Psi(\mathcal{C})$ is a left ideal of  $\mathbb{F}_q[D_{2n}]$. We will
equate $\mathcal{C}$ with $\Psi(\mathcal{C})$ in this paper.

\par
  There have been many research results
on codes as two-sided ideals and left ideals in a finite group algebra over finite fields. For example, Dutra et al \cite{Dutra2009Semisimple} investigated codes that are two-sided ideals in a semisimple finite group algebra
$\mathbb{F}_q[G]$,
and given a criterion to decide if these ideals are all the minimal two-sided ideals of $\mathbb{F}_q[G]$
when $G$ is a dihedral group. Brochero Mart\'{i}nez \cite{Mart2015Structure} showed all central irreducible
idempotents and their Wedderburn decomposition of the
semisimple dihedral group algebra $\mathbb{F}_q[D_{2n}]$ when every divisor
of $n$ divides $q-1$. Moreover, we gave a system theory for left $D_{2n}$-codes over finite fields $\mathbb{F}_q$ in \cite{Cao2016Concatenated} where ${\rm gcd}(q,n)=1$, and obtained a complete description for
left $D_{2n}$-codes over Galois rings ${\rm GR}(p^2,m)$ in \cite{Cao2018DM} were ${\rm gcd}(p,n)=1$.

\par
One of the most studied open questions
in coding theory is to ask whether there is an
extremal doubly-even binary self-dual
codes of length a multiple of $8$. There are still many problems worth studying in this field.
For example,

\noindent
  \textsf{For which $k$ does there exists a doubly-even self-dual
binary $[24k,12k,4k + 4]$ code} (Open Question 7.7 in \cite{Dougherty2013Open})?

\par
  In this paper, we provide a new way to construct binary self-dual
$[8m,4m]$-codes which is different from the methods used in
\cite{Mcloughlin2012A}, \cite{Dougherty2016Constructions}, \cite{Dougherty2018Group} and \cite{Dutra2009Semisimple}.
Specifically, we give an explicit construction and enumeration for
all distinct self-dual binary left $D_{8m}$-codes.
In future work,
we will try to determine extremal self-dual
binary $[8m,4m]$-codes among these codes.

\vskip 3mm \noindent
 \begin{notation}\label{no1.1}
In this paper, let $\mathbb{F}_2=\{0,1\}$ be a binary field and $m=2^{\lambda_0}m_0$, where $m_0$ and $\lambda_0$ are nonnegative integers such that $m_0$ is odd.
Then $8m=2\cdot 4m$ where
$$4m=2^\lambda\cdot m_0 \ {\rm and} \ \lambda=\lambda_0+2\geq 2.$$
\end{notation}

\par
    The present paper is organized as follows. In section II,
we introduce necessary notations and
give an explicit representation and enumeration for all distinct self-dual binary left $D_{8m}$-codes by Theorem \ref{th4.3} which is the main result of this paper. In Section III, we give recursive algorithms to solve
the problems in the construction of self-dual binary left $D_{8m}$-codes and obtain a clear formula to count the number of all these self-dual codes.
In Section IV, we list all distinct self-dual binary left $D_{8m}$-codes for
$m=1,3,6,7$. Among these codes, we obtain extremal
self-dual binary codes with parameters $[8,4,4]$, $[24,12,8]$, $[48,24,12]$, $[56,28,12]$, respectively.
In Section V, we give a detailed proof for Theorem \ref{th4.3} by four
subsections: Give a concatenated structure for every binary left $D_{8m}$-code;
Provide a representation and enumeration for all distinct binary left $D_{8m}$-codes;
Determine the dual code for each binary left $D_{8m}$-code; Prove Theorem \ref{th4.3}.
Section VI concludes the paper.

\section{Self-dual binary left $D_{8m}$-codes}

In this section, we introduce the necessary notations and
known results first.
Then we give an explicit representation and enumeration for all distinct self-dual binary left $D_{8m}$-codes.

\par
   For any nonzero polynomial $g(x)=\sum_{i=0}^da_ix^i\in \mathbb{F}_2[x]$ of degree $d$, the \textit{reciprocal polynomial}
of $g(x)$ is defined by
$$g^\ast(x)=(g(x))^\ast=x^dg(x^{-1})=a_d+a_{d-1}x+\ldots+a_0x^d,$$
and $g(x)$ is said to be \textit{self-reciprocal}
if $g^\ast(x)=g(x)$.

\par
  As $m_0$ is an odd positive integer, we have that
$$x^{m_0}-1=\prod_{i=0}^{r}f_i(x),$$
where $f_0(x),f_1(x),\ldots,f_r(x)$ are pairwise coprime irreducible polynomials in $\mathbb{F}_2[x]$ such that
\begin{itemize}
\item
  $r=\rho+2\epsilon$ for some nonnegative integers $\rho$ and $\epsilon$.

\item
   $f_0(x)=x+1$ with degree $d_0=1$.

\item
  $f_i(x)$ is self-reciprocal and of degree $d_i\geq 2$ for all $i=1,\ldots,\rho$.

\item
  $f_{\rho+j}(x)$ is not self-reciprocal, $f_{\rho+j}^\ast(x)=f_{\rho+j+\epsilon}(x)$
and ${\rm deg}(f_{\rho+j}(x))={\rm deg}(f_{\rho+j+\epsilon}(x))=d_{\rho+j}$ for all $j=1,\ldots,\epsilon$.
\end{itemize}

\vskip 2mm\noindent
It is clear that $m_0=\sum_{i=0}^rd_i$  and
$x^{4m}-1=\prod_{i=0}^{r}f_i(x)^{2^\lambda}.$
This implies
$4m=2^\lambda\sum_{i=0}^rd_i$.
In this paper, we denote:

\vskip 2mm\par
 $\bullet$ $\mathcal{A}=\mathbb{F}_2[x]/\langle x^{4m}-1\rangle$  where we regard elements of $\mathcal{A}$ as polynomials
in $\mathbb{F}_2[x]$ of degree $<4m$ and the arithmetic is done modulo $x^{4m}-1$.

\vskip 2mm\par
 $\bullet$ $A_i=\mathbb{F}_2[x]/\langle f_i(x)^{2^\lambda}\rangle$ where we regard elements of $A_i$ as polynomials
in $\mathbb{F}_2[x]$ of degree $<2^\lambda d_i$ and the arithmetic is done modulo $f_i(x)^{2^\lambda}$, for all $i=0,1,\ldots,r$.

\vskip 2mm \par
  Moreover, for any integers $i$ and $s$: $0\leq i\leq \rho$ and $1\leq s\leq 2^\lambda$, We adopt the following notation
in this paper.

\vskip 2mm\par
 $\bullet$  Let $\mathbb{F}_2[x]/\langle f_i(x)^s\rangle=\{\sum_{j=0}^{sd_i-1}a_jx^j\mid a_j\in \mathbb{F}_2,  j=0,1$, $\ldots,sd_i-1\}$
  in which the arithmetic is done modulo $f_i(x)^s$.

\vskip 2mm\par
 $\bullet$   Let $\mathcal{W}_i^{(s)}$ be the set of elements $w(x)$ in $\mathbb{F}_2[x]/\langle f_i(x)^s\rangle$ satisfying
$$w(x)w(x^{-1})\equiv 1 \ ({\rm mod} \ f_i(x)^s),$$
where $x^{-1}=x^{4m-1}$ (mod $f_i(x)^s$).

\vskip 2mm\par
 The rings $A_i$ ($0\leq i\leq r$) play important roles in this paper and their structures can be found in many dispersive literature.

\vskip 3mm \noindent
  \begin{lemma}\label{la2.3}  (cf. \cite{Cao2015Repeated} Example 2.1)
   \textit{Using the notations above, denote
$\langle f_i(x)^s\rangle=f_i(x)^sA_i$ for any $1\leq s\leq 2^\lambda$ and $0\leq i\leq r$. Then we have the following conclusions}.

\vskip 2mm\par
  (i) \textit{$A_i$ is a finite chain ring, $\langle f_i(x)\rangle$ is the unique
maximal ideal of $A_i$, the nilpotency index of $f_i(x)$ is $2^\lambda$ and $A/\langle f_i(x)\rangle\cong \mathbb{F}_2[x]/\langle f_i(x)\rangle\cong \mathbb{F}_{2^{d_i}}$}.

\vskip 2mm\par
  (ii) \textit{Let ${\cal T}_i=\{\sum_{j=0}^{d_i-1}t_jx^j\mid t_0,t_1,\ldots,t_{{d_i}-1}\in \mathbb{F}_2\}\subset A_i$. Then every element of $A_i$ has a unique $f_i(x)$-expansion}:
$$\sum_{j=0}^{2^\lambda-1}a_j(x)f_i(x)^j,
\ {\rm where} \ a_j(x)\in {\cal T}_i, \ \forall j=0,1,\ldots,2^\lambda-1.$$

\par
  (iii) \textit{We can regard elements of $A_i/\langle f_i(x)^s\rangle$ as the same as the
  ring $\mathbb{F}_2[x]/\langle f_i(x)^s\rangle$. Hence
  $|A_i/\langle f_i(x)^s\rangle|=2^{sd_i}$}.
\end{lemma}
\vskip 3mm\par
   For each $0\leq i\leq r$, denote
$F_i(x)=\frac{x^{m_0}-1}{f_i(x)}\in \mathbb{F}_2[x]$. Then $F_i(x)$ and $f_i(x)$ are coprime polynomials. Hence there are
polynomials $u_i(x),v_i(x)\in \mathbb{F}_2[x]$ such that
$$u_i(x)F_i(x)+v_i(x)f_i(x)=1.$$
This implies $F_i(x^{2^\lambda})=\frac{x^{4m}-1}{f_i(x^{2^\lambda})}=\frac{x^{4m}-1}{f_i(x)^{2^\lambda}}$ and
\begin{equation}\label{eq3}
u_i(x^{2^\lambda})F_i(x^{2^\lambda})+v_i(x^{2^\lambda})f_i(x)^{2^\lambda}=1.
\end{equation}
In the rest of this paper, let $\varepsilon_i(x)\in \mathcal{A}$ satisfying
\begin{equation}\label{eq4}
 \varepsilon_i(x)\equiv u_i(x^{2^\lambda})F_i(x^{2^\lambda}) \  ({\rm mod} \ x^{4m}-1).
\end{equation}

 \par
  From classical ring theory and the Chinese Remainder Theorem, we deduce the following
lemma (cf. \cite{Cao2015Repeated} Lemma 3.2).

\vskip 3mm \noindent
\begin{lemma}\label{la2.4}
  (i) \textit{$\sum_{i=0}^{r}\varepsilon_i(x)=1$, $\varepsilon_i(x)^2=\varepsilon_i(x)$ and $\varepsilon_i(x)\varepsilon_j(x)=0$ for all $0\leq i\neq j\leq r$ in the ring ${\cal A}$}.

\vskip 2mm\par
  (ii) \textit{${\cal A}=\bigoplus_{i=0}^{r}{\cal A}_i$, where ${\cal A}_i=\varepsilon_i(x){\cal A}$
with $\varepsilon_i(x)$ as its multiplicative identity. Moreover, this decomposition is a ring direct
sum in that ${\cal A}_i{\cal A}_j=\{0\}$ for all $0\leq i\neq j\leq r$}.

\vskip 2mm\par
  (iii) \textit{For each $0\leq i\leq r$, the map
$$\varphi_i: a(x)\mapsto \varepsilon_i(x)a(x) \ ({\rm mod} \ x^{4m}-1), \ \forall a(x)\in A_i$$
is an isomorphism of rings from $A_i$ onto ${\cal A}_i$}.

\vskip 2mm\par
  (iv) \textit{For any $a_i(x)\in A_i$, $0\leq i\leq r$, define}
$$\varphi: (a_0(x),\ldots,a_{r}(x))\mapsto \sum_{i=0}^{r}\varepsilon_i(x)a_i(x) \ ({\rm mod} \ x^{4m}-1).$$
\textit{Then $\varphi$ is a ring isomorphism from the direct product ring $A_0\times A_1\times\ldots\times A_{r}$ onto $\mathcal{A}$}.
\end{lemma}
\vskip 3mm \par
   As usual, we equate each vector $(a_0,a_1,a_2,\ldots,a_{4m-1})\in \mathbb{F}_2^{4m}$ with
$a_0+a_1x+\ldots+a_{4m-1}x^{4m-1}\in \mathcal{A}$. Then binary cyclic codes of length $4m$ are identified
with ideals of the ring $\mathcal{A}$. In particular, we have
the following properties for the ideal ${\cal A}_i=\varepsilon_i(x){\cal A}$ of ${\cal A}$.

\vskip 3mm \noindent
\begin{corollary}\label{co2.5}
 \textit{Let $0\leq i\leq r$. Then}

\vskip 2mm\par
   (i) \textit{${\cal A}_i$ is a binary cyclic code of length $4m$ with
parity check polynomial $f_i(x)^{2^\lambda}$ and generating idempotent $\varepsilon_i(x)$}.

\vskip 2mm\par
   (ii) \textit{As a binary linear code of length $4m$,
$$\{\varepsilon_i(x),x\varepsilon_i(x), x^2\varepsilon_i(x),\ldots,x^{2^\lambda d_i-1}\varepsilon_i(x)\}$$
is a basis of ${\cal A}_i$. Hence
${\rm dim}_{\mathbb{F}_2}({\cal A}_i)=2^\lambda d_i$}.
\end{corollary}

\vskip 3mm \noindent
   \begin{IEEEproof}
   (ii) Since $A_i$ is an $\mathbb{F}_2$-linear space with a basis $\{1,x,\ldots,x^{2^\lambda d_i-1}\}$, by Lemma \ref{la2.4} (iii)
we see that $\{\varepsilon_i(x)$, $x\varepsilon_i(x), \ldots,x^{2^\lambda d_i-1}\varepsilon_i(x)\}$
is an $\mathbb{F}_2$-basis of ${\cal A}_i$.
\end{IEEEproof}

\vskip 3mm\par
  Now, let $C_i$ be a linear code of length $2$ over $A_i$, i.e. $C_i$ is an $A_i$-submodule of $A_i^2=\{(b_0(x),b_1(x))\mid b_0(x),b_1(x)\in A_i\}$.
For each $\xi=(b_0(x),b_1(x))\in A_i^2$, we denote by
$${\rm w}_{H}^{(A_i)}(\xi)=|\{j\mid b_j(x)\neq 0 \ {\rm in} \ A_i, \ j=0,1\}|$$
 the Hamming weight of $\xi$ and define the
 minimum Hamming distance of $C_i$ as
$$d_{H}^{(A_i)}(C_i)={\rm min}\{{\rm w}_{H}^{(A_i)}(\xi) \mid \xi\neq 0, \ \xi\in C_i\}.$$
   As a natural generalization of the concept for concatenated codes over finite field (cf. \cite{Sendrier1998On}, Definition 2.1), using the notations of Lemma \ref{la2.4} (iii) we define the
\textit{concatenated code} $\mathcal{A}_i\Box_{\varphi_i}C_i$ of the inner code $\mathcal{A}_i$ and the outer code $C_i$ by
\begin{eqnarray*}
\mathcal{A}_i\Box_{\varphi_i}C_i&=&\{(\varphi_i(\xi_0),\varphi_i(\xi_1))\mid (\xi_0,\xi_1)\in C_i\}\\
 &=&\{(\varepsilon_i(x)\xi_0,\varepsilon_i(x)\xi_1)\mid (\xi_0,\xi_1)\in C_i\}\\
 &\subseteq & \mathcal{A}_i^2.
\end{eqnarray*}
By Lemma \ref{la2.4} (iii), we conclude that $\mathcal{A}_i\Box_{\varphi_i}C_i$ is a binary quasi-cyclic code of length $8m$ and index $2$
and the number of codewords is equal to $|\mathcal{A}_i\Box_{\varphi_i}C_i|=|C_i|$. This implies
$${\rm dim}_{\mathbb{F}_2}(\mathcal{A}_i\Box_{\varphi_i}C_i)={\rm log}_2|C_i|,$$
and the minimum Hamming distance of
$\mathcal{A}_i\Box_{\varphi_i}C_i$ satisfies
$$d_{H}^{(\mathbb{F}_2)}(\mathcal{A}_i\Box_{\varphi_i}C_i)\geq d_{H}^{(\mathbb{F}_2)}(\mathcal{A}_i)\cdot d_{H}^{(A_i)}(C_i),$$
where $d_{H}^{(\mathbb{F}_2)}(\mathcal{A}_i)$ is the minimum Hamming weight of $\mathcal{A}_i$ as a binary linear code of
length $4m$.

\par
  For the end of this section, we list all distinct self-dual binary left $D_{8m}$-codes by the following theorem.

\vskip 3mm \noindent
\begin{theorem}\label{th4.3}
   \textit{All distinct self-dual binary left $D_{8m}$-codes are given by
$${\cal C}=\bigoplus_{i=0}^{r}(\mathcal{A}_i\Box_{\varphi_i}C_i),$$
where $C_i$ is a linear code of length $2$ over $A_i$ with
a generator matrix $G_i$ given by the following cases}:

\vskip 2mm\par
  ($\dag$) \textit{Let $0\leq i\leq \rho$. Then $G_i$ is given by one of the following
$1+\sum_{j=1}^{2^{\lambda-1}}|\mathcal{W}_i^{(2j)}|$ matrices}:

\vskip 2mm\par
  ($\dag$-1) \textit{$|\mathcal{W}_i^{(2^\lambda)}|$ matrices}:
$$G_i=(1,a(x)), \ {\rm where} \ a(x)\in \mathcal{W}_i^{(2^\lambda)}.$$

\par
  ($\dag$-2) \textit{$1$ matrix: $G_i=f_i(x)^{2^{\lambda-1}}I_2$}.

\vskip 2mm\par
  ($\dag$-3) \textit{$\sum_{j=1}^{2^{\lambda-1}-1}|\mathcal{W}_i^{(2j)}|$ matrices}:
$$G_i=\left(\begin{array}{cc}f_i(x)^k & f_i(x)^k c(x) \cr 0 & f_i(x)^{2^\lambda-k}\end{array}\right),$$
\textit{where $c(x)\in \mathcal{W}_i^{(2^\lambda-2k)}$ and $1\leq k\leq 2^{\lambda-1}-1$}.

\vskip 2mm\par
  ($\ddag$) \textit{Let $\rho+1\leq i\leq \rho+\epsilon$ and denote}
$$\Omega_{(\lambda,d_i)}=1+2^{(2^\lambda-1) d_i}+2^{2^\lambda d_i}
+(2^{d_i}+1)\sum_{l=0}^{2^\lambda-3}(2^\lambda-2-l)2^{ld_i}.$$
 \textit{Then
the pair $(G_i,G_{i+\epsilon})$ is given by one of the following
$\Omega_{(\lambda,d_i)}$ pairs of matrices}.

\vskip 2mm\par
  ($\ddag$-1) \textit{$2^{2^\lambda d_i}$ pairs}:
$$G_i=(1,a(x)) \ {\rm and} \ G_{i+\epsilon}=(a(x^{-1}),1),$$
\textit{where $a(x)\in A_i$}.

\vskip 2mm\par
  ($\ddag$-2) \textit{$2^{(2^\lambda-1) d_i}$ pairs}:
$$G_i=(f_i(x)b(x),1) \ {\rm and} \ G_{i+\epsilon}=(1,f_{i+\epsilon}(x)\cdot x^{-d_i}b(x^{-1})),$$
\textit{where $b(x)\in \mathbb{F}_2[x]/\langle f_i(x)^{2^\lambda-1}\rangle$}.

\vskip 2mm\par
  ($\ddag$-3) \textit{$1$ pair}:
$$G_i=f_i(x)^{2^{\lambda-1}}I_2 \ {\rm and} \ G_{i+\epsilon}=f_{i+\epsilon}(x)^{2^{\lambda-1}}I_2.$$

\vskip 2mm\par
  ($\ddag$-4) \textit{$\sum_{k=1}^{2^\lambda-2}\sum_{j=1}^{2^\lambda-k-1}2^{jd_i}$ pairs}:

\par
 $G_i=\left(\begin{array}{cc} f_i(x)^{2^\lambda-k-j} & f_i(x)^{2^\lambda-k-j}c(x)\cr
 0 & f_{i}(x)^{2^\lambda-k}\end{array}\right)$  \textit{and}
$$G_{i+\epsilon}=\left(\begin{array}{cc}f_{i+\epsilon}(x)^{2^\lambda-k-j}c(x^{-1}) & f_{i+\epsilon}(x)^{2^\lambda-k-j} \cr  f_{i+\epsilon}(x)^{2^\lambda-k} & 0\end{array}\right),$$

\noindent
\textit{where $c(x)\in \mathbb{F}_2[x]/\langle f_i(x)^j\rangle$,
$1\leq j\leq 2^\lambda-k-1$ and $1\leq k\leq 2^\lambda-2$}.

\vskip 2mm\par
  ($\ddag$-5) \textit{$\sum_{k=1}^{2^\lambda-2}\sum_{j=1}^{2^\lambda-k-1}2^{(j-1)d_i}$ pairs}:

\par
 $G_i=\left(\begin{array}{cc} f_i(x)^{2^\lambda-k-j}c(x) & f_i(x)^{2^\lambda-k-j}\cr
f_{i}(x)^{2^\lambda-k} & 0\end{array}\right)$  \textit{and}
 $$G_{i+\epsilon}=\left(\begin{array}{cc}f_{i+\epsilon}(x)^{2^\lambda-k-j}
 & f_{i+\epsilon}(x)^{2^\lambda-k-j}c(x^{-1})
 \cr 0 & f_{i+\epsilon}(x)^{2^\lambda-k} \end{array}\right),$$

\noindent
\textit{where $c(x)\in f_i(x)(\mathbb{F}_2[x]/\langle f_i(x)^j\rangle)$,
$1\leq j\leq 2^\lambda-k-1$ and $1\leq k\leq 2^\lambda-2$}.

\vskip 2mm\par
  \textit{Then the number of self-dual binary left $D_{8m}$-codes is}
$$\left(\prod_{i=0}^\rho(1+\sum_{j=1}^{2^{\lambda-1}}|\mathcal{W}_i^{(2j)}|)\right)\cdot \left(\prod_{i=\rho+1}^{\rho+\epsilon}\Omega_{(\lambda,d_i)}\right).$$
\end{theorem}

\vskip 3mm\par
   To make it convenient for readers, we put a detailed proof for this theorem in Section V.

   Now, in order to list self-dual binary left $D_{8m}$-codes by use of Theorem \ref{th4.3},
we have to solve the following problem:

\noindent
\textsf{Give an efficient algorithm to determine the set $\mathcal{W}_i^{(s)}$ for
all integers $s=2j$ and $i$: $1\leq j\leq 2^{\lambda-1}$ and $0\leq i\leq \rho$}.


\section{Recursive algorithm to calculate $\mathcal{W}_i^{(s)}$}
\noindent
   In this section, we consider how to calculate the sets $\mathcal{W}_i^{(s)}$
defined in Section II, for any $1\leq s\leq 2^\lambda$ and
$0\leq i\leq \rho$.

\begin{center}
  $\diamondsuit$ $1\leq i\leq \rho$
\end{center}

\par
  Let $1\leq i\leq \rho$. Then $d_i$ is even since
$f_i(x)$ is self-reciprocal and irreducible.
When $s=1$, we have the following lemma.

\vskip 3mm
\begin{lemma}\label{la5.3}
 (cf. \cite{Cao2016Concatenated} Theorem \ref{th3.2} and its proof)
\textit{Using the notations above, let $\zeta_i(x)$ be a primitive element of the finite field $\mathbb{F}_2[x]/\langle f_i(x)\rangle$,
i.e., the multiplicative order of $\zeta_i(x)$ modulo $f_i(x)$ is equal to $2^{d_i}-1$. Then}

\vskip 2mm\par
  (i) \textit{$a(x^{-1})\equiv a(x)^{2^{\frac{d_i}{2}}}$ $({\rm mod} \ f_i(x))$, for any $a(x)\in \mathbb{F}_2[x]$}.

\vskip 2mm\par
  (ii) \textit{The congruence equation $a(x)a(x^{-1})\equiv 1$  $({\rm mod} \ f_i(x))$ has exactly $1+2^{\frac{d_i}{2}}$ solutions
in $\mathbb{F}_2[x]/\langle f_i(x)\rangle$: $a_1(x)=\zeta_i(x)^{(2^{\frac{d_i}{2}}-1)l}$ for $l=0,1,\ldots,2^{\frac{d_i}{2}}$. Hence}
$$\mathcal{W}_i^{(1)}=\{\zeta_i(x)^{(2^{\frac{d_i}{2}}-1)l}\mid l=0,1,\ldots,2^{\frac{d_i}{2}}\}
\ ({\rm mod} \ f_i(x))$$
\textit{and $|\mathcal{W}_i^{(1)}|=2^{\frac{d_i}{2}}+1$}.
\end{lemma}
\vskip 3mm \par
  In order to express $\mathcal{W}_i^{(s)}$ precisely for $s=2,3,\ldots,2^\lambda$, we introduce the following notations:

\par
  $\bullet$ $\mathcal{K}_i=\mathbb{F}_2[x]/\langle f_i(x)\rangle$, i.e.,
$$\mathcal{K}_i=\left\{\sum_{j=0}^{d_i-1}c_jx^j\mid c_j\in \mathbb{F}_2, \
0\leq j\leq d_i-1\right\}$$
in which the arithmetic is done modulo $f_i(x)$. Then $\mathcal{K}_i$ is a finite field of $2^{d_i}$ elements.

\par
  $\bullet$ $\mathcal{F}_i=\{\xi\in \mathcal{K}_i\mid \xi^{2^{\frac{d_i}{2}}}=\xi\}$. Then $\mathcal{F}_i$ is the unique subfield of $\mathcal{K}_i$ with $2^{\frac{d_i}{2}}$ elements. Precisely, we have
$$\mathcal{F}_i=\{0\}\cup \left\{\zeta_i(x)^{(2^{\frac{d_i}{2}}+1)t}\mid t=0,1,\ldots,2^{\frac{d_i}{2}}-2\right\}.$$

\par
  $\bullet$  ${\rm Tr}_{\mathcal{K}_i/\mathcal{F}_i}$ is the trace function from $\mathcal{K}_i$ onto $\mathcal{F}_i$, i.e.,
$${\rm Tr}_{\mathcal{K}_i/\mathcal{F}_i}(\alpha)=\alpha^{2^{\frac{d_i}{2}}}+\alpha=a(x)^{2^{\frac{d_i}{2}}}+a(x) \ ({\rm mod} \ f_i(x)),$$
for any $\alpha=a(x)\in \mathcal{K}_i$.
Then from \cite{Wan2006Lectures}
Corollary 7.17 (i), we deduce the following conclusion.

\vskip 3mm
\begin{lemma}\label{la5.4}
\textit{For any $\gamma\in \mathcal{F}_i$,
we denote
$${\rm Tr}_{\mathcal{K}_i/\mathcal{F}_i}^{-1}(\gamma)=\{\beta\in \mathcal{K}_i\mid {\rm Tr}_{\mathcal{K}_i/\mathcal{F}_i}(\beta)=\beta^{2^{\frac{d_i}{2}}}+\beta=\gamma\}.$$
Then $|{\rm Tr}_{\mathcal{K}_i/\mathcal{F}_i}^{-1}(\gamma)|=2^{\frac{d_i}{2}}$}.
\end{lemma}

\vskip 3mm
\par
  We calculate $\mathcal{W}_i^{(s)}$ recursively by the following theorem.

\vskip 3mm \noindent
\begin{theorem}\label{th5.5}
\textit{Let $1\leq i\leq \rho$ and $2\leq s\leq 2^\lambda$. Assume $\mathcal{W}_i^{(s-1)}$ has been determined.
Then $\mathcal{W}_i^{(s)}$
can be determined by the following three steps}:

\vskip 2mm \noindent
  Step 1. \textit{Choose $a_{s-1}(x)\in \mathcal{W}_i^{(s-1)}$ arbitrary, and calculate}
$$b_{s-1}(x)=\frac{a_{s-1}(x)a_{s-1}(x^{4m-1})-1}{f_i(x)^{s-1}} \ ({\rm mod} \ f_i(x)).$$
\textit{Then $x^{(s-1)\frac{d_i}{2}}b_{s-1}(x)\in \mathcal{F}_i$}.

\vskip 2mm \noindent
  Step 2. \textit{Find $\beta(x)\in \mathcal{K}_i=\mathbb{F}_2[x]/\langle f_i(x)\rangle$ such that
$\beta(x)\in {\rm Tr}_{\mathcal{K}_i/\mathcal{F}_i}^{-1}\left(x^{(s-1)\frac{d_i}{2}}b_{s-1}(x)\right)$, i.e.,}
$$\beta(x)^{2^{\frac{d_i}{2}}}+\beta(x)\equiv x^{(s-1)\frac{d_i}{2}}b_{s-1}(x)  \ ({\rm mod} \ f_i(x)).$$

\noindent
  Step 3. \textit{Set $a_1(x)=a_{s-1}(x)$ $({\rm mod} \ f_i(x))$, and calculate}
$$z(x)=x^{4m-(s-1)\frac{d_i}{2}}a_1(x)\beta(x) \ ({\rm mod} \ f_i(x)).$$
\textit{Then $a_s(x)=a_{s-1}(x)+z(x)f_i(x)^{s-1}\in \mathcal{W}_i^{(s)}$}.

\vskip 2mm \par
  \textit{Therefore, the number of elements in $\mathcal{W}_i^{(s)}$ is equal to} $$|\mathcal{W}_i^{(s)}|=(2^{\frac{d_i}{2}}+1)2^{(s-1)\frac{d_i}{2}}.$$
\end{theorem}

\vskip 3mm
\begin{IEEEproof}
Let $a_{s-1}(x)\in \mathcal{W}_i^{(s-1)}$. Then by the definition of $\mathcal{W}_i^{(s-1)}$ and $x^{-1}=x^{4m-1}$ in $A_i=\mathbb{F}_2[x]/\langle f_i(x)^{2^\lambda}\rangle$, we have
$$a_{s-1}(x)a_{s-1}(x^{4m-1})\equiv 1 \ ({\rm mod} \ f_i(x)^{s-1}).$$
This implies $f_i(x)^{s-1}|\left(a_{s-1}(x)a_{s-1}(x^{4m-1})-1\right)$ in $\mathbb{F}_2[x]$, and so $\frac{a_{s-1}(x)a_{s-1}(x^{4m-1})-1}{f_i(x)^{s-1}}$ $\in \mathbb{F}_2[x]$. Hence the polynomial $b_{s-1}(x)$
calculated by Step 1 belong to $\mathcal{K}_i$. As polynomials in $\mathbb{F}_2[x]$,
by $f_i(x^{-1})=x^{-d_i}f_i^\ast(x)=x^{-d_i}f_i(x)$ we have that
\begin{eqnarray*}
b_{s-1}(x^{-1})&\equiv&\frac{a_{s-1}(x^{-1})a_{s-1}(x^{-(-1)})-1}{f_i(x^{-1})^{s-1}}\\
   &=&\frac{a_{s-1}(x^{-1})a_{s-1}(x)-1}{x^{-(s-1)d_i}f_i(x)^{s-1}}\\
&=&x^{(s-1)d_i}\frac{a_{s-1}(x^{-(4m-1)})a_{s-1}(x)-1}{f_i(x)^{s-1}}\\
&\equiv& x^{(s-1)d_i}b_{s-1}(x) \ ({\rm mod} \ f_i(x)).
\end{eqnarray*}
From this and by Lemma \ref{la5.3}(i), we deduce that
\begin{eqnarray*}
&&\left(x^{(s-1)\frac{d_i}{2}}b_{s-1}(x)\right)^{2^{\frac{d_i}{2}}}=(x^{-1})^{(s-1)\frac{d_i}{2}}b_{s-1}(x^{-1})\\
     &=& x^{-(s-1)\frac{d_i}{2}}\cdot x^{(s-1)d_i}b_{s-1}(x)
  = x^{(s-1)\frac{d_i}{2}}b_{s-1}(x)
\end{eqnarray*}
in $\mathcal{K}_i$. This implies $x^{(s-1)\frac{d_i}{2}}b_{s-1}(x)\in \mathcal{F}_i$, since $\mathcal{F}_i$ is a subfield of $\mathcal{K}_i$
and $|\mathcal{F}_i|=2^{\frac{d_i}{2}}$.

\par
  Let $a_s(x)=a_{s-1}(x)+z(x)f_i(x)^{s-1}$ where $z(x)\in\mathcal{K}_i$. Then by
$a_{s-1}(x)a_{s-1}(x^{4m-1})\equiv 1+b_{s-1}(x)f_i(x)^{s-1}$ (mod $f_i(x)$), it follows that
\begin{eqnarray*}
&&a_s(x)a_s(x^{-1})\\
  &=&\left(a_{s-1}(x)+z(x)f_i(x)^{s-1}\right)\\
  &&\cdot \left(a_{s-1}(x^{-1})+z(x^{-1})f_i(x^{-1})^{s-1}\right)\\
 &=&\left(a_{s-1}(x)x^{-(s-1)d_i}z(\frac{1}{x})+a_{s-1}(\frac{1}{x})z(x)\right)f_i(x)^{s-1}\\
 &&+a_{s-1}(x)a_{s-1}(x^{4m-1})\\
 &&+z(x)z(x^{-1})x^{-(s-1)d_i}f_i(x)^{s-1}\cdot f_i(x)^{s}\\
 &\equiv& 1+f_i(x)^{s-1}\left(a_{s-1}(x)x^{-(s-1)d_i}z(x^{-1})\right. \\
 && \left.+a_{s-1}(x^{-1})z(x)+b_{s-1}(x)\right) \ ({\rm mod} \ f_i(x)^{s}).
\end{eqnarray*}
From this we deduce that $a_s(x)\in \mathcal{W}_i^{(s)}$, i.e. $a_s(x)a_s(x^{-1})\equiv 1$ (mod $f_i(x)^{s}$), if and only if
$$a_{s-1}(x)x^{-(s-1)d_i}z(x^{-1})+a_{s-1}(x^{-1})z(x)+b_{s-1}(x)\equiv 0$$
(mod $f_i(x)$).
Then by Lemma \ref{la5.3} (i),
$a_{1}(x)\equiv a_{s-1}(x)$ (mod $f_i(x)$) and $a_1(x)\in \mathcal{W}_i^{(1)}$, we see that the latter
condition is equivalent to that $z(x)\in\mathcal{K}_i$ satisfying the following condition
\begin{equation}\label{eq10}
\varrho(x)\equiv 0  \ ({\rm mod} \ f_i(x)),
\end{equation}
where
$$\varrho(x)=x^{-(s-1)d_i}a_{1}(x)z(x)^{2^{\frac{d_i}{2}}}+a_{1}(x)^{2^{\frac{d_i}{2}}}z(x)+b_{s-1}(x).$$
Furthermore, by $a_1(x)\in \mathcal{W}_i^{(1)}$ and Lemma \ref{la5.3} we have
$$a_{1}(x)=a_{1}(x)^{-2^{\frac{d_i}{2}}}, \
a_{1}(x)^{2^{\frac{d_i}{2}}}=a_1(x^{-1}), \ x^{-1}=x^{2^{\frac{d_i}{2}}}.$$
Multiplying $x^{(s-1)\frac{d_i}{2}}$ on both sides of Equation (\ref{eq10}), we obtain
$$(\frac{x^{(s-1)\frac{d_i}{2}}z(x)}{a_1(x)})^{2^{\frac{d_i}{2}}}+\frac{x^{(s-1)\frac{d_i}{2}}z(x)}{a_1(x)}
+x^{(s-1)\frac{d_i}{2}}b_{s-1}(x)=0.$$
Now, set $\beta(x)=\frac{x^{(s-1)\frac{d_i}{2}}z(x)}{a_1(x)}\in \mathcal{K}_i$. Then we have $z(x)=x^{4m-(s-1)\frac{d_i}{2}}a_1(x)\beta(x)$ (mod $f_i(x)$), where
$\beta(x)$ satisfies
$$\beta(x)^{2^{\frac{d_i}{2}}}+\beta(x)=x^{(s-1)\frac{d_i}{2}}b_{s-1}(x),$$
i.e.,
$\beta(x)\in {\rm Tr}_{\mathcal{K}_i/\mathcal{F}_i}^{-1}\left(x^{(s-1)\frac{d_i}{2}}b_{s-1}(x)\right)$.

\par
  Finally, by Lemmas \ref{la5.3} and \ref{la5.4} it follows that $|\mathcal{W}_i^{(s)}|=|\mathcal{W}_i^{(s-1)}|\cdot 2^{\frac{d_i}{2}}=\ldots=
(2^{\frac{d_i}{2}}+1) 2^{(s-1)\frac{d_i}{2}}$.
\end{IEEEproof}

\vskip 3mm\par
  Therefore, for any $1\leq i\leq \rho$ we have
$$1+\sum_{j=1}^{2^{\lambda-1}}|\mathcal{W}_{i}^{(2j)}|
=1+(2^{\frac{d_i}{2}}+1)\sum_{j=1}^{2^{\lambda-1}}2^{(2j-1)\frac{d_i}{2}}.$$
From this and by $2^{(2j-1)\frac{d_i}{2}}=2^{\frac{d_i}{2}}(2^{d_i})^{j-1}$, we deduce that
$$\bullet \ 1+\sum_{j=1}^{2^{\lambda-1}}|\mathcal{W}_{i}^{(2j)}|
=1+(2^{d_i}+2^{\frac{d_i}{2}})\frac{2^{2^{\lambda-1}d_i}-1}{2^{d_i}-1}
\ {\rm if} \ 1\leq i\leq \rho.$$

\begin{center}
  $\diamondsuit\diamondsuit$ $i=0$
\end{center}

\par
   Now,
  let $i=0$. Then $f_0(x)=x+1$,
$$A_0=\mathbb{F}_2[x]/\langle (x+1)^{2^\lambda}\rangle=\mathbb{F}_2[x]/\langle x^{2^\lambda}+1\rangle$$
and $x^{-1}=x^{2^\lambda-1}$ in $A_0$. So we have $A_0/\langle x+1\rangle=\mathbb{F}_2$ and

\par
  $\bullet$ $\mathcal{W}_0^{(1)}=\{1\}.$

\par
When $s\geq 2$, $\mathcal{W}_0^{(s)}$ can be calculated recursively by the following theorem.

\vskip 3mm
\begin{theorem}\label{th5.1}
 \textit{Let $2\leq s\leq 2^\lambda$. For any $a_{s-1}(x)\in \mathcal{W}_0^{(s-1)}$, calculate
\begin{equation}\label{eq8}
b_{s-1}\equiv \frac{a_{s-1}(x)a_{s-1}(x^{2^\lambda-1})-1}{(x+1)^{s-1}} \ ({\rm mod} \ x+1).
\end{equation}
Then we have one of the following cases}:

\vskip 2mm\par
  (i) \textit{If $b_{s-1}=0$, then $a_{s-1}(x), a_{s-1}(x)+(x+1)^{s-1}\in \mathcal{W}_0^{(s)}$}.

\vskip 2mm\par
  (ii) \textit{If $b_{s-1}=1$, then $a_{s-1}(x)\not\in \mathcal{W}_0^{(s)}$ and $a_{s-1}(x)+(x+1)^{s-1}\not\in \mathcal{W}_0^{(s)}$}.
\end{theorem}

\vskip 3mm
\begin{IEEEproof}
 By Equation (\ref{eq8}), as a polynomial in $\mathbb{F}_2[x]$ we have
$$a_{s-1}(x)a_{s-1}(x^{2^\lambda-1})=1+b_{s-1}(x+1)^{s-1} \  ({\rm mod} \ (x+1)^s).$$
Denote $a(x)=a_{s-1}(x)+c_{s-1}(x+1)^{s-1}\in A_0/\langle (x+1)^s\rangle$ and $u(x)=1+\sum_{t=1}^{2^\lambda-1}(x+1)^t$,
where $c_{s-1}\in \mathbb{F}_2$. Then
$$x\cdot u(x)=(1+(x+1))u(x)=1+(x+1)^{2^\lambda}=1 \ {\rm in} \ A_0.$$
This implies $x^{2^\lambda-1}=x^{-1}=u(x)$ in $A_0$. Therefore, we have
$$(x^{2^\lambda-1}+1)^{s-1}=((x+1)u(x))^{s-1}=(x+1)^{s-1}u(x)^{s-1}$$
and $u(x)\equiv u(1)=1$ (mod $x+1$). Hence as a polynomial in $\mathbb{F}_2[x]$, we have
\begin{eqnarray*}
&&a(x)a(x^{2^\lambda-1})\\
&=&\left(a_{s-1}(x)+c_{s-1}(x+1)^{s-1}\right)\\
  &&\cdot\left(a_{s-1}(x^{2^\lambda-1})+c_{s-1}(x^{2^\lambda-1}+1)^{s-1}\right)\\
  &=&a_{s-1}(x)a_{s-1}(x^{2^\lambda-1})\\
  &&+c_{s-1}^2(x+1)^{s}\cdot (x+1)^{s-2}u(x)^{s-1}\\
  &&+(x+1)^{s-1}\cdot c_{s-1}(a_{s-1}(x)u(x)^{s-1}+a_{s-1}(x^{2^\lambda-1})).
\end{eqnarray*}
From this and by
\begin{eqnarray*}
 && a_{s-1}(x)u(x)^{s-1}+a_{s-1}(x^{2^\lambda-1})\\
 &\equiv & a_{s-1}(1)u(1)^{s-1}+a_{s-1}(1)=0 \ ({\rm mod} \ x+1),
\end{eqnarray*}
we deduce that
\begin{equation}\label{eq9}
a(x)a(x^{2^\lambda-1})\equiv a_{s-1}(x)a_{s-1}(x^{2^\lambda-1}) \ ({\rm mod} \ (x+1)^{s}).
\end{equation}

\par
  (i) Let $b_{s-1}=0$. Then
$$a_{s-1}(x)a_{s-1}(x^{-1})\equiv a_{s-1}(x)a_{s-1}(x^{2^\lambda-1})\equiv 1$$
(mod $(x+1)^s$). This implies
$a(x)a(x^{-1})=1$ in $A_0/\langle (x+1)^s\rangle$ by Equation (\ref{eq9}).
 Hence $a(x)=a_{s-1}(x)+c_{s-1}(x+1)^{s-1}\in \mathcal{W}_0^{(s)}$
for all $c_{s-1}\in \mathbb{F}_2$.

\par
  (ii) Let $b_{s-1}=1$. As a polynomial in $\mathbb{F}_2[x]$, we have that $a_{s-1}(x)a_{s-1}(x^{2^\lambda-1})\equiv 1+(x+1)^{s-1}$ (mod $(x+1)^{s}$). Then by Equation (\ref{eq9}), in the ring $A_0/\langle (x+1)^s\rangle$ we have
$$a(x)a(x^{-1})=a_{s-1}(x)a_{s-1}(x^{-1})=1+(x+1)^{s-1}\neq 1.$$
Hence $a(x)=a_{s-1}(x)+c_{s-1}(x+1)^{s-1}\not\in \mathcal{W}_0^{(s)}$
for any $c_{s-1}\in\mathbb{F}_2$.
\end{IEEEproof}

\vskip 3mm
\begin{example}\label{ex5.5}
   Let $\lambda=3$. We calculate $\mathcal{W}_0^{(s)}$ for all $s=1,2,3,\ldots,8$.
For $a_1(x)\in \mathcal{W}_0^{(1)}=\{1\}$, we have $a_1(x)=1$. As $b_1=\frac{a_1(x)a_1(x^7)-1}{x+1}=0$,
by Theorem \ref{th5.1} we obtain

\par
 $\bullet$ $\mathcal{W}_0^{(2)}=\{1,1+(x+1)\}=\{1,x\}$.

\par
  For any $a_2(x)\in \mathcal{W}_0^{(2)}$, we have $b_2=\frac{a_2(x)a_2(x^7)-1}{(x+1)^2}\equiv 0$ (mod $x+1$). Then
by Theorem \ref{th5.1} it follows that

\par
 $\bullet$ $\mathcal{W}_0^{(3)}=\{1,1+(x+1)^2,x,x+(x+1)^2\}=\{1,x,x^2$, $1+x+x^2\}$.

\par
  For any $a_3(x)\in \mathcal{W}_0^{(3)}$, by a direct calculation we get $b_3=\frac{a_3(x)a_3(x^7)-1}{(x+1)^3}\equiv 0$ (mod $x+1$). By Theorem \ref{th5.1} we obtain
\begin{eqnarray*}
\bullet \ \mathcal{W}_0^{(4)}&=&\{a_3(x),a_3(x)+(x+1)^3\mid a_3(x)\in \mathcal{W}_0^{(3)}\}\\
  &=&\{1,x,x^2,x^3,1+x+x^2,1+x^2+x^3,\\
  &&x+x^2+x^3,1+x+x^3\}.
\end{eqnarray*}

\par
   For any $a_4(x)\in \{1,x,x^2,x^3\}\subset\mathcal{W}_0^{(4)}$, it is clear that $b_4=\frac{a_4(x)a_4(x^{7})-1}{(x+1)^4}\equiv 0$ (mod $x+1$), but for
any $a_4(x)\in \{1+x+x^2,1+x^2+x^3,x+x^2+x^3,1+x+x^3\}\subset\mathcal{W}_0^{(4)}$ we
have $b_4=\frac{a_4(x)a_4(x^7)-1}{(x+1)^4}\equiv 1$ (mod $x+1$).
Hence by Theorem \ref{th5.1} we obtain
\begin{eqnarray*}
\bullet \ \mathcal{W}_0^{(5)}&=&\{a_4(x),a_4(x)+(x+1)^4\\
  && \ \mid a_4(x)\in \{1,x,x^2,x^3\}\}\\
  &=&\{1,x,x^2,x^3,x^4,1+x^2+x^4,\\
  &&1+x+x^4,1+x^3+x^4\}.
\end{eqnarray*}

\noindent
Using a similar method, we get the following calculations:

\par
  $\bullet$ $\mathcal{W}_0^{(6)}=\{a_5(x), a_5(x)+(x+1)^5\mid a_5(x)\in \mathcal{W}_0^{(5)}\}$ with $|\mathcal{W}_0^{(6)}|=2|\mathcal{W}_0^{(5)}|=16=2^4.$
\begin{eqnarray*}
  \bullet \ \mathcal{W}_0^{(7)}&=&\{a_5(x), a_5(x)+(x+1)^6 \\
  && \ \ \mid a_5(x)\in \{1,x,x^2,x^3,x^4,1+x^2+x^4\}\} \\
  &&\cup\{a_6(x), a_6(x)+(x+1)^6 \\
  && \ \ \mid a_6(x)=a_5(x)+(x+1)^5, \\
  && \ \ a_5(x)\in \{1+x+x^4,1+x^3+x^4\}\}
\end{eqnarray*}
with $|\mathcal{W}_0^{(7)}|=2|\mathcal{W}_0^{(5)}|=2^4.$

\par
  $\bullet$ $\mathcal{W}_0^{(8)}=\{a_7(x), a_7(x)+(x+1)^8\mid a_7(x)\in \mathcal{W}_0^{(7)}\}$ with $|\mathcal{W}_0^{(8)}|=2|\mathcal{W}_0^{(7)}|=2^5.$
\end{example}

\vskip 3mm\par
 Moreover, we have $|\mathcal{W}_0^{(9)}|=2^5$ and $|\mathcal{W}_0^{(10)}|=|\mathcal{W}_0^{(11)}|=2^6$ (for $\lambda\geq 4$).

\par
  As stated above, we conclude the following conclusion.

\vskip 3mm
\begin{corollary}\label{co5.6}
\textit{Using Notation 1.1 in Section I,
The number of self-dual binary
left $D_{2^{\lambda+1}m_0}$-codes is}
$$\omega_\lambda
\left(\prod_{i=1}^\rho(1+(2^{d_i}+2^{\frac{d_i}{2}})\frac{2^{2^{\lambda-1}d_i}-1}{2^{d_i}-1})\right)
\left(\prod_{i=\rho+1}^{\rho+\epsilon}\Omega_{(\lambda,d_i)}\right),$$

\noindent
\textit{where
$\omega_\lambda=1+\sum_{j=1}^{2^{\lambda-1}}|\mathcal{W}_0^{(2j)}|$ and
$$\Omega_{(\lambda,d_i)}=1+2^{(2^\lambda-1) d_i}+2^{2^\lambda d_i}
+(2^{d_i}+1)\sum_{l=0}^{2^\lambda-3}(2^\lambda-2-l)2^{ld_i}.$$
In particular, we have the following formulas}:

\vskip 2mm\par
  $\diamond$ \textit{When $\lambda=2$, The number of self-dual binary
left $D_{8m_0}$-codes is}
$$11\left(\prod_{i=1}^\rho\left(1+(2^{d_i}+2^{\frac{d_i}{2}})(2^{d_i}+1)\right)\right)
\left(\prod_{i=\rho+1}^{\rho+\epsilon}\Omega_{(2,d_i)}\right),$$
\textit{where} $\Omega_{(2,d_i)}=1+2^{3d_i}+2^{4d_i}
+(2^{d_i}+1)(2+2^{d_i})$.

\vskip 2mm\par
  $\diamond$ \textit{When $\lambda=3$, The number of self-dual binary
left $D_{16m_0}$-codes is}
$$59\left(\prod_{i=1}^\rho(1+(2^{d_i}+2^{\frac{d_i}{2}})\frac{2^{4d_i}-1}{2^{d_i}-1})\right)
\left(\prod_{i=\rho+1}^{\rho+\epsilon}\Omega_{(3,d_i)}\right),$$
\textit{where} $\Omega_{(3,d_i)}=1+2^{7 d_i}+2^{8d_i}
+(2^{d_i}+1)\sum_{l=0}^{5}(6-l)2^{ld_i}$.
\end{corollary}

\vskip 3mm\par
  Let $SLD(\lambda,m_0)$ be the number of all self-dual binary
left $D_{2^{\lambda+1}m_0}$-codes. Then we have the following table:
{\small
\begin{center}
\begin{tabular}{ll|l}\hline
 $2^{\lambda+1}m_0$ & $(\lambda,m_0))$ & $SLD(\lambda,m_0)$  \\ \hline
 $8$  & $(2,1)$ & $11$\\
 $24$ & $(2,3)$ & $11\cdot 31=341$\\
 $40$ & $(2,5)$ & $11\cdot 341=3751$\\
 $56$ & $(2,7)$ & $11\cdot 4699=51689$\\
 $72$ & $(2,9)$ & $11\cdot 31\cdot 4681=1596221$\\
 $16$  & $(3,1)$ & $59$\\
 $48$ & $(3,3)$ & $59\cdot 511=30149$\\
 $80$ & $(3,5)$ & $59\cdot 87381=5155479$\\
 $112$ & $(3,7)$ & $59\cdot 19259551=1136313509$\\
 $144$ & $(3,9)$ & $59\cdot 511\cdot 19173961=578075750189$\\
\hline
\end{tabular}
\end{center}  }

\vskip 3mm\noindent
  {\bf Conjecture} $|\mathcal{W}_0^{(s)}|=2^{1+\lfloor \frac{s}{2}\rfloor}=2\cdot 2^{\lfloor \frac{s}{2}\rfloor}$ for any integer $s\geq 4$. Then
$$\omega_\lambda=1+\sum_{j=1}^{2^{\lambda-1}}|\mathcal{W}_0^{(2j)}|=1+2+2\sum_{j=2}^{2^{\lambda-1}}2^j
=3+8(2^{2^{\lambda-1}-1}-1).$$
This conjecture has been proven to hold for $\lambda=2,3$.

\section{Self-dual binary left $D_{8m}$-codes for $m=1,3, 6, 7$}
 In this section, we describe in detail how to list explicitly all distinct
 self-dual binary left $D_{8}$-codes, left $D_{24}$-codes, left $D_{48}$-codes and left $D_{56}$-codes, respectively.

 \vskip 3mm
\begin{example}\label{ex6.1}
 We consider self-dual binary left $D_8$-codes. In this case, we have $\lambda=2$,
$m=1$ and $x^4-1=(x+1)^4$ in $\mathbb{F}_2[x]$. Then $r=\rho=\epsilon=0$ and $\varepsilon_0(x)=1$. By Theorem \ref{th4.3}, all distinct self-dual binary left $D_8$-codes
are given by: $C_0$, where $C_0$ is a linear code of length $2$ over the finite chain ring $A_0=\mathbb{F}_2[x]/\langle (x+1)^4\rangle$ with one and only one of the following $11$ matrices as its generator matrix:

  (i) \textit{$G_0=(1,a(x))$, $a(x)\in \mathcal{W}_0^{(4)}$};

\par
  (ii) $G_0=(x+1)^{2}I_2$;

\par
  (iii) \textit{$G_0=\left(\begin{array}{cc}(x+1) & (x+1)c(x) \cr 0 & (x+1)^{3}\end{array}\right)$,
$c(x)\in \mathcal{W}_0^{(2)}$},

\noindent
 where $\mathcal{W}_0^{(4)}$ and $\mathcal{W}_0^{(2)}$ are given in Example \ref{ex5.5}.

\par
  Each self-dual binary left $D_8$-code is a self-dual binary $[8,4,d]$-codes where
$d$ is the minimal Hamming distance of $C_0$. Precisely, we have the following
table:
\begin{center}
\begin{tabular}{llll}\hline
case &  $G_0$  &  $[8,4,d]$    \\ \hline
(i)  & $(1,a(x))$, $a(x)\in \{1,x,x^2,x^3\}$&  $[8,4,2]$ \\
     & $(1,a(x))$, $a(x)\in \mathcal{W}_0^{(4)}\setminus \{1,x,x^2,x^3\}$ &  \textbf{[8,4,4]} \\
(ii) & $(x+1)^{2}I_2$ &  $[8,4,2]$ \\
(iii)& $\left(\begin{array}{cc}x+1 & (x+1)c(x) \cr 0 & (x+1)^{3}\end{array}\right)$, $c(x)\in \{1,x\}$ &  \textbf{[8,4,4]} \\
 \hline
\end{tabular}
\end{center}
\end{example}

\vskip 3mm \noindent
\begin{example}\label{ex6.2}
 We consider binary left $D_{24}$-codes. In this case, $24=2\cdot 12$, $12=4\cdot 3$ and $x^3-1=f_0(x)f_1(x)$ where $f_0(x)=x+1$
and $f_1(x)=x^2+x+1$. Hence $\rho=r=1$, $\epsilon=0$,
$d_0=1$, $d_1=2$ and $x^{12}-1=f_0(x)^4f_1(x)^4.$

\par
   Let $\mathcal{A}=\mathbb{F}_2[x]/\langle x^{12}-1\rangle$, $A_0=\mathbb{F}_2[x]/\langle (1+x)^4\rangle$,
$A_1=\mathbb{F}_2[x]/\langle (1+x+x^2)^4\rangle$,
$\varepsilon_0(x)=1+x^4+x^{8}$ and $\varepsilon_1(x)=x^4+x^{8}$.
   By Theorem \ref{th4.3}, all $341$ self-dual binary left $D_{24}$-codes are given by:
$\mathcal{C}=(\mathcal{A}_0\Box_{\varphi_0}C_0)\oplus(\mathcal{A}_1\Box_{\varphi_1}C_1)$,
where

\par
  $\diamondsuit$  $\mathcal{A}_0$ is a binary cyclic code of length $12$ with idempotent generator $\varepsilon_0(x)$
and parity check polynomial $1+x^4$.

\par
  $\varphi_0$ is an isomorphism of rings from $A_0$ onto $\mathcal{A}_0$ defined by $\varphi_0(a(x))=\varepsilon_0(x)a(x)$
(mod $x^{12}+1$) for all $a(x)\in A_0$.

\par
  $C_0$ is one of the $11$ linear codes of length $2$ over $A_0$ with
a generator matrix $G_0$ given in Example \ref{ex6.1}.

\par
  $\diamondsuit$  $\mathcal{A}_1$ is a binary cyclic code of length $12$ with idempotent generator $\varepsilon_1(x)$
and parity check polynomial $1+x^4+x^{8}$.

\par
  $\varphi_1$ is an isomorphism of rings from $A_1$ onto $\mathcal{A}_1$ defined by $\varphi_1(b(x))=\varepsilon_1(x)b(x)$
(mod $x^{12}+1$) for all $b(x)\in A_1$.

\par
  $C_1$ is a linear code of length $2$ over $A_1$ with a generator matrix $G_1$ given by
one of the following $31$ matrices:

\vskip 2mm\par
  (i) $G_1=(1,a(x))$, $a(x)\in \mathcal{W}_1^{(4)}$;

\vskip 2mm\par
  (ii) $G_1=f_1(x)^{2}I_2$ where $f_1(x)=x^2+x+1$;

\vskip 2mm\par
  (iii) $G_1=\left(\begin{array}{cc} f_1(x) & f_1(x)c(x) \cr 0 & f_1(x)^3\end{array}\right)$,
$c(x)\in \mathcal{W}_1^{(2)}$,

\noindent
 where $\mathcal{W}_1^{(2)}$ and $\mathcal{W}_1^{(4)}$ are given in Appendix B.

\par
   Among the $341$ codes listed above, we have $24$ self-dual binary $[24,12,8]$-codes with the same weight distribution
enumerator $1+759X^8+2576X^{12}+759X^{16}+X^{24}$: $$\mathcal{C}=(\mathcal{A}_0\Box_{\varphi_0}C_0)\oplus(\mathcal{A}_1\Box_{\varphi_1}C_1),$$
where the generator matrix $G_0$ of $C_0$ and the generator matrix $G_1$ of $C_1$ are given by
one of the two cases:

\par
  ($\dag$) $G_0=(1,a_0(x))$ with $a_0(x)\in\{1+x+x^2,1+x^2+x^3\}$;

\par
   $G_1=(1,a_1(x))$ with $a_1(x)\in \{x+x^4+x^7, x+x^3$ $+x^5+x^6,
  x+x^2+x^6+x^7, 1+x+x^3+x^5,x^2+x^3+x^5$ $+x^7,1+x^3+x^4+x^5+x^7\}$.

\par
  ($\ddag$) $G_0=(1,a_0(x))$ with $a_0(x)\in\{1+x+x^3,x+x^2+x^3\}$;

\par
  $G_1=(1,a_1(x))$ with $a_1(x)\in \{1+x^3+x^6,x+x^2$ $+x^4+x^6,
1+x+x^5+x^6, x^2+x^4+x^6+x^7,1+x^2+x^4$ $+x^5,1+x^2+x^3+x^4+x^7\}$.
\end{example}

\par
 The above $24$ self-dual binary $[24,12,8]$-codes are extremal and permutation equivalent.

\vskip 3mm \noindent
\begin{example}\label{ex6.3}
We consider binary left $D_{48}$-codes. In this case, $48=2\cdot 24$, $24=8\cdot 3$ and $x^3-1=f_0(x)f_1(x)$ where
$f_0(x)=x+1$ and $f_1(x)=x^2+x+1$. Hence $r=1$, $d_0=1$, $d_1=2$ and
$x^{24}-1=f_0(x)^8f_1(x)^8.$
Let $\mathcal{A}=\mathbb{F}_2[x]/\langle x^{24}-1\rangle$, $A_0=\mathbb{F}_2[x]/\langle (1+x)^8\rangle$,
$A_1=\mathbb{F}_2[x]/\langle (1+x+x^2)^8\rangle$,
 $$\varepsilon_0(x)=1+x^8+x^{16} \ {\rm and} \
 \varepsilon_1(x)=x^8+x^{16}.$$

\par
   By Theorem \ref{th4.3}, all $30149$ self-dual binary left $D_{48}$-codes are given by:
$\mathcal{C}=(\mathcal{A}_0\Box_{\varphi_0}C_0)\oplus(\mathcal{A}_1\Box_{\varphi_1}C_1),$
here for $i=0,1$:

\par
  $\diamond$  $\mathcal{A}_i$ is a binary cyclic code of length $24$ with idempotent generator $\varepsilon_i(x)$
and parity check polynomial $f_i(x)^8$.

\par
 $\diamond$ $\varphi_i$ is an isomorphism of rings from $A_i$ onto $\mathcal{A}_i$ defined by $\varphi_i(a(x))=\varepsilon_i(x)a(x)$
(mod $x^{24}+1$) for all $a(x)\in A_i$.

\par
 $\diamond$ $C_0$ is a linear code of length $2$ over $A_0$ with a generator matrix $G_0$ given by
one of the following three cases:

\vskip 2mm\par
  (i) $G_0=(1,a(x))$, $a(x)\in \mathcal{W}_0^{(8)}$;

\vskip 2mm\par
  (ii) $G_0=(x+1)^{4}I_2$;

\vskip 2mm\par
  (iii) $G_0=\left(\begin{array}{cc}(x+1)^k & (x+1)^k c(x) \cr 0 & (x+1)^{8-k}\end{array}\right)$,
$c(x)\in \mathcal{W}_0^{(8-2k)}$ and $1\leq k\leq 3$,

\noindent
  where $\mathcal{W}_0^{(8-2k)}$ is given in Example \ref{ex5.5} for all $k=0,1,2,3$.

\par
 $\diamond$ $C_1$ is a linear code of length $2$ over $A_1$ with a generator matrix $G_1$ given by
one of the following three cases:

\vskip 2mm\par
  (i) $G_1=(1,b(x))$, $b(x)\in \mathcal{W}_1^{(8)}$;

\vskip 2mm\par
  (ii) $G_1=f_1(x)^{4}I_2$ where $f_1(x)=x^2+x+1$;

\vskip 2mm\par
  (iii) $G_1=\left(\begin{array}{cc}f_1(x)^k & f_1(x)^k c(x) \cr 0 & f_1(x)^{8-k}\end{array}\right)$,
$c(x)\in \mathcal{W}_1^{(8-2k)}$ and $1\leq k\leq 3$,

\noindent
  where $\mathcal{W}_1^{(2)}$ and $\mathcal{W}_1^{(4)}$ are given in Appendix B, $\mathcal{W}_1^{(6)}$ and $\mathcal{W}_1^{(8)}$ can be calculated easily by use of the algorithm in Theorem \ref{th5.5}. Here we omit the
calculation results to save spaces, since $|\mathcal{W}_1^{(6)}|=96$ and $|\mathcal{W}_1^{(8)}|=384$.

\par
   Among the $30149$ codes listed above, we have $192$ doubly-even self-dual binary $[48,24,12]$-codes: $$\mathcal{C}=(\mathcal{A}_0\Box_{\varphi_0}C_0)\oplus(\mathcal{A}_1\Box_{\varphi_1}C_1),$$
where $C_0$ has the generator matrix $G_0=(1,a(x))$, $C_1$ has the generator matrix $G_1=(1,b(x))$ and the pair
$(a(x),b(x))$ is given by Appendix C of this paper. These $192$ self-dual binary $[48,24,12]$-codes are extremal and permutation equivalent.
\end{example}

\vskip 3mm \noindent
\begin{example}\label{ex6.4}
We consider binary left $D_{56}$-codes. In this case, $56=2\cdot 28$, $28=4\cdot 7$, $\lambda=2$ and $x^{7}-1
=f_0(x)f_1(x)f_2(x)$ where $f_0(x)=x+1$, $f_1(x)=x^3+x+1$ and $f_2(x)=x^3+x^2+1=f^\ast_2(x)$.
Hence $r=2$, $\rho=0$, $\epsilon=1$, $d_0=1$, $d_1=d_2=3$ and
$x^{28}-1=f_0(x)^4f_1(x)^4f_2(x)^4.$
Let $\mathcal{A}=\mathbb{F}_2[x]/\langle x^{28}-1\rangle$, $A_0=\mathbb{F}_2[x]/\langle (1+x)^4\rangle$,
$A_i=\mathbb{F}_2[x]/\langle f_i(x)^4\rangle$ for $i=1,2$,
$\varepsilon_0(x)=x^{24}+x^{20}+x^{16}+x^{12}+x^8+x^4+1$,
 $$\varepsilon_1(x)=x^{16}+x^8+x^4+1 \ {\rm and} \
 \varepsilon_2(x)=x^{24}+x^{20}+x^{12}+1.$$

\par
   By Theorem \ref{th4.3}, all $51689$ self-dual binary left $D_{56}$-codes are given by
$\mathcal{C}=(\mathcal{A}_0\Box_{\varphi_0}C_0)\oplus(\mathcal{A}_1\Box_{\varphi_1}C_1)\oplus(\mathcal{A}_2\Box_{\varphi_2}C_2),$
where for $i=0,1,2$ we have the following:

\par
  $\diamond$  $\mathcal{A}_i$ is a binary cyclic code of length $24$ with idempotent generator $\varepsilon_i(x)$
and parity check polynomial $f_i(x)^4$.

\par
 $\diamond$ $\varphi_i$ is an isomorphism of rings from $A_i$ onto $\mathcal{A}_i$ defined by $\varphi_i(a(x))=\varepsilon_i(x)a(x)$
(mod $x^{28}+1$) for all $a(x)\in A_i$.

\par
 $\diamond$ $C_0$ is one of the $11$ linear codes of length $2$ over $A_0$ with
a generator matrix $G_0$ given in Example \ref{ex6.1}.

\par
 $\diamond$ $C_i$ is a linear code of length $2$ over $A_i$ with a generator matrix $G_i$ for $i=1,2$, and
the pair $(G_1,G_2)$ is given by
one of the following $4699$ pairs of matrices where $x^{-1}=x^{27}$ (mod $f_2(x)^4$):

\vskip 2mm\par
  (i) $2^{12}=4096$ pairs:
$$G_1=(1,\eta(x)) \ {\rm and} \ G_{2}=(\eta(x^{-1}),1),
\ {\rm where} \ \eta(x)\in A_1.$$

\par
  (ii) $2^{9}=512$ pairs:
$$G_1=(f_1(x)b(x),1) \ {\rm and} \ G_{2}=(1,f_{2}(x)\cdot x^{-3}b(x^{-1})),$$
where $b(x)\in A_1/\langle f_1(x)^{3}\rangle$.

\vskip 2mm\par
  (iii) $1$ pair:
$G_1=f_1(x)^{2}I_2 \ {\rm and} \ G_{2}=f_{2}(x)^{2}I_2.$

\vskip 2mm\par
  (iv) $80$ pairs:

\noindent
  {\bf 1.} $G_1=\left(\begin{array}{cc} f_1(x)^{2} & f_1(x)^{2}c(x)\cr
 0 & f_{1}(x)^{3}\end{array}\right)$ where $c(x)\in A_i/\langle f_i(x)\rangle$,
 and
 $G_{2}=\left(\begin{array}{cc}f_{2}(x)^{2}c(x^{-1}) & f_{2}(x)^{2} \cr  f_{2}(x)^{3} & 0\end{array}\right)$;

\noindent
  {\bf 2.} $G_1=\left(\begin{array}{cc} f_1(x) & f_1(x)c(x)\cr
 0 & f_{1}(x)^{3}\end{array}\right)$ where $c(x)\in A_i/\langle f_i(x)^2\rangle$,
 and
 $G_{2}=\left(\begin{array}{cc}f_{2}(x)c(x^{-1}) & f_{2}(x) \cr  f_{2}(x)^{3} & 0\end{array}\right)$;

\noindent
  {\bf 3.} $G_1=\left(\begin{array}{cc} f_1(x) & f_1(x)c(x)\cr
 0 & f_{1}(x)^{2}\end{array}\right)$ where $c(x)\in A_i/\langle f_i(x)\rangle$,
 and
 $G_{2}=\left(\begin{array}{cc}f_{2}(x)c(x^{-1}) & f_{2}(x) \cr  f_{2}(x)^{2} & 0\end{array}\right)$.

\vskip 2mm\par
  (v) \textit{$10$ pairs}:

\noindent
{\bf 1.} $G_1=\left(\begin{array}{cc} 0 & f_1(x)^{2}\cr
f_{1}(x)^{3} & 0\end{array}\right)$,
$G_{2}=\left(\begin{array}{cc}f_{2}(x)^{2}
 & 0
 \cr 0 & f_{2}(x)^{3} \end{array}\right)$;

\noindent
{\bf 2.}
 $G_1=\left(\begin{array}{cc} f_1(x)c(x) & f_1(x)\cr
f_{1}(x)^{2} & 0\end{array}\right)$ and
$$G_{2}=\left(\begin{array}{cc}f_{2}(x)
 & f_{2}(x)c(x^{-1})
 \cr 0 & f_{2}(x)^{3} \end{array}\right),$$
where $c(x)=f_1(x)a(x)$ and $a(x)$ is a polynomial
in $\mathbb{F}_2[x]$ of degree less than $3$;

\noindent
{\bf 3.} $G_1=\left(\begin{array}{cc} 0 & f_1(x)\cr
f_{1}(x)^{2} & 0\end{array}\right)$,
$G_{2}=\left(\begin{array}{cc}f_{2}(x)
 & 0
 \cr 0 & f_{2}(x)^{2} \end{array}\right)$.

\vskip 3mm\par
   Among $51689$ self-dual binary left $D_{56}$-codes, we have the following $728$ doubly-even self-dual binary $[56,28,12]$-codes: $$\mathcal{C}=(\mathcal{A}_0\Box_{\varphi_0}C_0)\oplus(\mathcal{A}_1\Box_{\varphi_1}C_1)
\oplus(\mathcal{A}_2\Box_{\varphi_2}C_2),$$
where $G_0=(1,a(x))$, $G_1=(1,\eta(x))$ and $G_2=(\eta(x^{-1}),1)$ is
a generator matrix of the code $C_0$, $C_1$ and $C_2$, respectively, and
the pairs $(a(x),\eta(x))$ of polynomials are
given in Appendix D of this paper explicitly. These $728$ self-dual binary $[56,28,12]$-codes are extremal and permutation equivalent.
\end{example}

\section{Proof of Theorem \ref{th4.3}}

\noindent
 In this section, we give a proof for Theorem \ref{th4.3}. It needs to be divided into four parts.

\subsection*{V.1 Concatenated structure of binary left $D_{8m}$-codes}
  In this subsection, we give a concatenated structure for every binary  left $D_{8m}$-code. As
$$D_{8m}=\langle x,y\mid x^{4m}=1, y^2=1, yxy=x^{-1}\rangle,$$
 $C^{(4m)}=\langle x\mid x^{4m}=1\rangle$
is a cyclic subgroup of $D_{8m}$ with order $4m$ generated by $x$. Hence
the group algebra $\mathbb{F}_2[C^{(4m)}]$ is equal to the residue class ring $\mathcal{A}
=\mathbb{F}_2[x]/\langle x^{4m}-1\rangle$.
This implies that $\mathcal{A}$ is a subring of $\mathbb{F}_2[D_{8m}]$ and
$$\mathbb{F}_2[D_{8m}]=\{\alpha(x)+\beta(x)y\mid \alpha(x), \beta(x)\in \mathcal{A}\}\ (y^2=1)$$
in which $y\alpha(x)=\alpha(x^{-1})y$ for all $\alpha(x)\in \mathcal{A}$. Now, we define a map
$\Theta: \mathcal{A}^2\rightarrow \mathbb{F}_2[D_{8m}]$ by
$$\Theta(\alpha(x), \beta(x))=\alpha(x)+\beta(x)y, \
\forall\alpha(x), \beta(x)\in \mathcal{A}.$$
Then one can easily verify that $\Theta$ is an $\mathcal{A}$-module isomorphism from
$\mathcal{A}^2$ onto $\mathbb{F}_2[D_{8m}]$.

\par
   Let $\mathcal{C}$ be a nonempty subset of $\mathbb{F}_2[D_{8m}]$. Then $\mathcal{C}$ is a left ideal of $\mathbb{F}_2[D_{8m}]$,
i.e. $\mathcal{C}$ is a binary left $D_{8m}$-code, if and only if $\Theta^{-1}(\mathcal{C})$ is an $\mathcal{A}$-submodule
of $\mathcal{A}^2$ and $y\xi\in \mathcal{C}$ for any $\xi=\alpha(x)+\beta(x)y\in \mathcal{C}$. From this and
by
$$y\xi=\alpha(x^{-1})y+\beta(x^{-1})y^2=\beta(x^{-1})+\alpha(x^{-1})y,$$
we deduce that
$$y\xi\in \mathcal{C} \Longleftrightarrow (\beta(x^{-1}),\alpha(x^{-1}))\in \Theta^{-1}(\mathcal{C}).$$
Let $\mathcal{C}^\prime=\Theta^{-1}(\mathcal{C})$. Then $\mathcal{C}=\Theta(\mathcal{C}^\prime)$.
Hence
$\mathcal{C}$ is a binary left $D_{8m}$-code if and only if there is a unique $\mathcal{A}$-submodule $\mathcal{C}^\prime$
of $\mathcal{A}^2$ satisfying the following condition:
$$
(\beta(x^{-1}),\alpha(x^{-1}))\in \mathcal{C}^\prime, \ \forall (\alpha(x),\beta(x))\in \mathcal{C}^\prime
$$
 such that $\Theta(\mathcal{C}^\prime)=\mathcal{C}$. We will equate $\mathcal{C}$ with $\mathcal{C}^\prime$ in this paper.

\vskip 3mm\noindent
\begin{theorem}\label{th2.6}
 \textit{Every binary left $D_{8m}$-code $\mathcal{C}$ can be uniquely decomposed as the
following}:
$$\mathcal{C}=\bigoplus_{i=0}^r\left(\mathcal{A}_i\Box_{\varphi_i}C_i\right)
=\sum_{i=0}^r\left(\mathcal{A}_i\Box_{\varphi_i}C_i\right),$$
\textit{where $C_i$, $0\leq i\leq r$, is a linear code of length $2$ over the finite chain ring $A_i$ satisfying
the following conditions}:

\vskip 2mm\par
   (i) \textit{If $0\leq i\leq \rho$, $C_i$ satisfies}
\begin{equation}\label{eq5}
(b_i(x^{-1}),a_i(x^{-1}))\in C_i, \ \forall (a_i(x),b_i(x))\in C_i.
\end{equation}

\vskip 2mm\par
   (ii) \textit{If $\rho+1\leq i\leq \rho+\epsilon$, the pair $(C_i,C_{i+\epsilon})$ of linear codes is given by}
$$C_{i+\epsilon}=\{(b_i(x^{-1}),a_i(x^{-1}))\mid (a_i(x),b_i(x))\in C_i\}\subseteq A_{i+\epsilon}^2,$$
\textit{where $C_i$ is an arbitrary
linear code of length $2$ over $A_i$}.

\vskip 2mm\par
\textit{Moreover, the number of codewords in $\mathcal{C}$ is $\prod_{i=0}^{r}|C_i|$}.
\end{theorem}
\vskip 3mm \noindent
\begin{IEEEproof}
    For any integer $i$, $0\leq i\leq r$, denote
$$\mu(i)=\left\{\begin{array}{ll} i, & {\rm when} \ 0\leq i\leq \rho; \cr
                      i+\epsilon, & {\rm when} \ \rho+1\leq i\leq \rho+\epsilon; \cr
                      i-\epsilon, & {\rm when} \ \rho+\epsilon+1\leq i\leq \rho+2\epsilon.\end{array}\right.$$
  Now, we claim that
\begin{equation}\label{eq6}
\varepsilon_i(x^{-1})=\varepsilon_{\mu(i)}(x) \ {\rm in} \ \mathcal{A},
\ {\rm where} \ x^{-1}=x^{4m-1}.
\end{equation}
In fact, by $f_{\mu(i)}(x)=f_i^\ast(x)=x^{d_i}f_i(x^{-1})$ we have
$$f_i(x^{-1})=x^{-d_i}f_{\mu(i)}(x) \ {\rm in} \ \mathcal{A}$$
and ${\rm deg}(f_{\mu(i)}(x))={\rm deg}(f_i(x))=d_i$.
From these, we deduce that
$f_i(x^{-1})^{2^\lambda}=x^{-2^\lambda d_i}f_{\mu(i)}(x)^{2^\lambda}$ and
\begin{eqnarray*}
F_i(x^{-2^\lambda})&=&\frac{x^{-4m}-1}{f_i(x^{-1})^{2^\lambda}}=\frac{x^{4m}+1}{x^{4m}f_i(x^{-1})^{2^\lambda}}\\
 &=& \frac{x^{4m}+1}{x^{4m-2^\lambda d_i}f_{\mu(i)}(x)^{2^\lambda}}\\
 &=&x^{2^\lambda d_i-4m}F_{\mu(i)}(x).
\end{eqnarray*}
By Equation (\ref{eq3}), it follows that
$$u_i(x^{-2^\lambda})F_i(x^{-2^\lambda})+v_i(x^{-2^\lambda})f_i(x^{-1})^{2^\lambda}=1,$$
where
\begin{eqnarray*}
v_i(x^{-2^\lambda})f_i(x^{-1})^{2^\lambda}
  &=&v_i(x^{-2^\lambda})x^{-2^\lambda d_i}f_{\mu(i)}(x)^{2^\lambda} \\
  &=&\widehat{v}_{\mu(i)}(x)f_{\mu(i)}(x)^{2^\lambda},
\end{eqnarray*}
\begin{eqnarray*}
u_i(x^{-2^\lambda})F_i(x^{-2^\lambda})
 &=& u_i(x^{-2^\lambda})x^{2^\lambda d_i-4m}F_{\mu(i)}(x^{2^\lambda}) \\
 &=&\widehat{u}_{\mu(i)}(x)F_{\mu(i)}(x^{2^\lambda}),
\end{eqnarray*}
and $\widehat{v}_{\mu(i)}(x), \widehat{u}_{\mu(i)}(x)\in \mathcal{A}$ satisfying:

\par
  $\widehat{v}_{\mu(i)}(x)=x^{4m-2^\lambda (d_i+{\rm deg}(v_i(x))}v_i^\ast(x^{2^\lambda})$;

\par
 $\widehat{u}_{\mu(i)}(x)=x^{2^\lambda d_i+4m-2^\lambda{\rm deg}(u_i(x))}u_i^\ast(x^{2^\lambda})$,

\noindent
respectively. Hence
$$\widehat{u}_{\mu(i)}(x)F_{\mu(i)}(x^{2^\lambda})+\widehat{v}_{\mu(i)}(x)f_{\mu(i)}(x)^{2^\lambda}=1.$$
From these and by Equation (\ref{eq4}), we deduce that
\begin{eqnarray*}
\varepsilon_i(x^{-1})&=&u_i(x^{-2^\lambda})F_i(x^{-2^\lambda})=\widehat{u}_{\mu(i)}(x)F_{\mu(i)}(x^{2^\lambda})\\
  &=&\varepsilon_{\mu(i)}(x).
\end{eqnarray*}
Therefore, Equation (\ref{eq6}) was proved.

\par
  Using the notations in Lemma \ref{la2.4} (iii) and (iv), for any $(\xi_{i0},\xi_{i1})\in A_i^2$, $0\leq i\leq r$, we define
\begin{eqnarray*}
&&\Phi((\xi_{00},\xi_{01}),(\xi_{10},\xi_{11}),\ldots,(\xi_{r0},\xi_{r1}))\\
  &=&\sum_{i=0}^{r+t}(\varphi_i(\xi_{i0}),\varphi_i(\xi_{i1}))\\
  &=&(\sum_{i=0}^{r}\varepsilon_i(x)\xi_{i0},\sum_{i=0}^{r}\varepsilon_i(x)\xi_{i1})
   \ ({\rm mod} \ x^{4m}-1).
\end{eqnarray*}
It is clear that $\Phi$ is an $\mathbb{F}_2[x]$-module isomorphism from $A_0^2\times A_1^2\times\ldots\times A^2_{r}$
onto $\mathcal{A}^2$. Now, let $\mathcal{C}$ be an $\mathcal{A}$-submodule of $\mathcal{A}^2$. Then
for each integer $i$, $0\leq i\leq r$, there is a unique $A_i$-submodule $C_i$ of $A_i^2$ such that
\begin{eqnarray*}
\mathcal{C}&=&\Phi(C_0\times C_1\times\ldots\times C_{r})\\
 &=&\{\sum_{i=0}^{r}(\varphi_i(\xi_{i0}),\varphi_i(\xi_{i1}))\mid (\xi_{i0},\xi_{i1})\in C_i, \ 0\leq i\leq r\}\\
 &=&\bigoplus_{i=0}^{r}\{(\varphi_i(\xi_{i0}),\varphi_i(\xi_{i1}))\mid (\xi_{i0},\xi_{i1})\in C_i\}\\
 &=&
 (\mathcal{A}_0\Box_{\varphi_0}C_0)\oplus (\mathcal{A}_1\Box_{\varphi_1}C_1)\oplus\ldots
 \oplus(\mathcal{A}_{r}\Box_{\varphi_{r}}C_{r}).
\end{eqnarray*}
It is obvious that $|\mathcal{C}|=\prod_{i=0}^{r}|C_i|$.

\par
  Moreover, for any integer $i$, $0\leq i\leq r$, and $(a_i(x),b_i(x))\in C_i$, let $(\alpha(x),\beta(x))\in \mathcal{C}$ where
$$\alpha(x)=\sum_{i=0}^{r}\varepsilon_i(x)a_i(x) \ {\rm and} \
\beta(x)=\sum_{i=0}^{r}\varepsilon_i(x)b_i(x).$$
By Equation (\ref{eq6}), in the ring $\mathcal{A}$ we have

\par
$\alpha(x^{-1})=\sum_{i=0}^{r}\varepsilon_i(x^{-1})a_i(x^{-1})=\sum_{i=0}^{r}\varepsilon_{\mu(i)}(x)a_i(x^{-1})$;

\par
 $\beta(x^{-1})=\sum_{i=0}^{r}\varepsilon_i(x^{-1})b_i(x^{-1})=\sum_{i=0}^{r}\varepsilon_{\mu(i)}(x)b_i(x^{-1})$.

\noindent
 These imply
\begin{eqnarray*}
&&(\beta(x^{-1}),\alpha(x^{-1}))\\
 &=&(\sum_{i=0}^{r}\varepsilon_{\mu(i)}(x)b_i(x^{-1}),\sum_{i=0}^{r}\varepsilon_{\mu(i)}(x)a_i(x^{-1}))\\
  &=&\Phi((b_0(x^{-1}),a_0(x^{-1})),\ldots,(b_{\rho}(x^{-1}),a_{\rho}(x^{-1})),\\
    &&(b_{\rho+\epsilon+1}(x^{-1}),a_{\rho+\epsilon+1}(x^{-1})),\\
    &&\ldots,(b_{\rho+2\epsilon}(x^{-1}),a_{\rho+2\epsilon}(x^{-1})),\\
    &&(b_{\rho+1}(x^{-1}),a_{\rho+1}(x^{-1})),\\
    &&\ldots,(b_{\rho+\epsilon}(x^{-1}),a_{\rho+\epsilon}(x^{-1}))).
\end{eqnarray*}
From this and by $\mathcal{C}=\Phi(C_0\times C_1\times\ldots\times C_{r})$,  we deduce that
$$(\beta(x^{-1}),\alpha(x^{-1}))\in \mathcal{C}
\Longleftrightarrow (b_i(x^{-1}),a_i(x^{-1}))\in C_{\mu(i)}, \ \forall i.$$
Therefore, we have one of the following two cases:

\par
  (i) Let $0\leq i\leq \rho$. In this case, we have $\mu(i)=i$ and hence
$C_i$ satisfies Condition (\ref{eq5}).

\par
  (ii) Let $\rho+1\leq i\leq \rho+\epsilon$. we have $\mu(i)=i+\epsilon$ and
$\mu(i+\epsilon)=i$. In this case, $C_i$ and $C_{i+\epsilon}$ satisfy
the above conditions if and only if
$C_{i+\epsilon}=\{(b_i(x^{-1}),a_i(x^{-1}))\mid (a_i(x),b_i(x))\in C_i\}$ and
$C_i$ is an arbitrary linear code over $A_i$ of length $2$.
\end{IEEEproof}

\vskip 3mm \par
   In the rest of this paper,  we call ${\cal C}=\bigoplus_{i=1}^{r}(\mathcal{A}_i\Box_{\varphi_i}C_i)$ the \textit{canonical form decomposition} of the binary left $D_{8m}$-code
${\cal C}$.

By Theorem \ref{th2.6}, in order to list all distinct binary left $D_{8m}$-codes
we just need to solve the following questions:

\par
  \textsl{Question 1}. Determine all linear codes of length $2$ over $A_i$ for
each $i=\rho+1,\ldots,\rho+\epsilon$.

\par
 \textsl{Question 2}. Determine all linear codes of length $2$ over $A_i$ satisfying Condition (\ref{eq5}) for
each $i=0,1,\ldots,\rho$.


\subsection*{V.2 Representation and enumeration for binary left $D_{8m}$-codes}
  In this subsection,
 we solve the two questions at the end of Subsection V.1
first. Then we obtain an explicit representation and enumeration for all distinct binary left $D_{8m}$-codes.

\par
  Let $0\leq i\leq r$ and $a(x)\in A_i=\mathbb{F}_2[x]/\langle f_i(x)^{2^\lambda}\rangle$. By Lemma \ref{la2.3}(ii), $a(x)$ has a unique $f_i(x)$-adic expansion:
$$
a(x)=\sum_{j=0}^{2^\lambda-1}f_i(x)^{j}a_{j}(x), \ a_j(x)\in \mathcal{T}_i.$$
If $a(x)\neq 0$, we define the \textit{$f_i(x)$-degree} of $a(x)$ as the least index $j$ for which $a_j(x)\neq 0$ and denote as $\|a(x)\|_{f_i(x)}=j$. If $\alpha(x)=0$
we define $\|a(x)\|_{f_i(x)}=2^\lambda$.

Furthermore, for any vector $\alpha=(a(x),b(x))\in A_i^2$, where $a(x),b(x)\in A_i$, we define the \textit{$f_i(x)$-degree} of $\alpha$
by
$$\|\alpha\|_{f_i(x)}={\rm min}\{\|a(x)\|_{f_i(x)}, \|b(x)\|_{f_i(x)}\}.$$
Let $1\leq j\leq 2^\lambda$ and $b(x)\in A_i/\langle f_i(x)^j\rangle=\mathbb{F}_2[x]/\langle f_i(x)^j\rangle$. By Lemma \ref{la2.3} (iii), $b(x)$
has a unique $f_i(x)$-expansion:
$$b(x)=b_0(x)+f_i(x) b_1(x)+\ldots+f_i(x)^{j-1}b_{j-1}(x)+\langle f_i(x)^j\rangle$$
where $b_l(x)\in \mathcal{T}_i$ for
all $l=0,1,\ldots,j-1$. So
$$|A_i/\langle f_i(x)^j\rangle|=|\mathcal{T}_i|^{j}=2^{jd_i}$$
(cf. \cite{Norton2000On} or  \cite{Hou2003On}). In the rest of this paper, we will identify the element $b(x)\in A_i/\langle f_i(x)^j\rangle$ with $\sum_{l=0}^{j-1}f_i(x)^l b_l(x)\in A_i$ and
regard $A_i/\langle f_i(x)^j\rangle$ as a subset of $A_i$.

\par
   Let $C_i$ be a linear code over $A_i$ of length $2$. By \cite{Norton2000On}
Definition 3.1, a matrix $G_i$ is called a \textit{generator matrix} for $C_i$ if the rows of $G_i$ span $C_i$ and none of them can be written as an $A_i$-linear combination of the other rows of $G_i$.

\par
 $\diamondsuit$ First, for Question 1 at the end of Subsection V.1 we have the following conclusion:

\vskip 3mm
\begin{lemma}\label{la2.2}
 (cf. \cite{Cao2015Repeated} Example 2.5) \textit{Using the notation above,
let $0\leq i\leq r$. Then the number of linear codes
over $A_i$ of length $2$ is equal to
$$\mathcal{L}_{(2,2^{d_i},2^\lambda)}=\sum_{0\leq k\leq 2^\lambda}(2k+1)2^{(2^\lambda-k)d_i}.$$
 Precisely, every linear code $C_i$ over
$A_i$ of length $2$ has one and only one of the following matrices $G_i$ as its generator matrices in standard form}:

\vskip 2mm \noindent
{\bf 1.} $G_i=(1,a(x))$, \textit{where} $a(x)\in A_i$.

\vskip 2mm \noindent
{\bf 2.} $G_i=(f_i(x)^k,f_i(x)^ka(x))$, \textit{where} $1\leq k\leq 2^\lambda-1$ \textit{and}  $a(x)\in A_i/\langle f_i(x)^{2^\lambda-k}\rangle$.

\vskip 2mm \noindent
{\bf 3.} $G_i=(f_i(x) b(x),1)$, \textit{where} $b(x)\in A_i/\langle f_i(x)^{2^\lambda-1}\rangle$.

\vskip 2mm \noindent
{\bf 4.} $G_i=(f_i(x)^{k+1}b(x),f_i(x)^k)$, \textit{where} $1\leq k\leq 2^\lambda-1$ \textit{and} $b(x)\in A_i/\langle f_i(x)^{2^\lambda-k-1}\rangle$.

\vskip 2mm \noindent
{\bf 5.} $G_i=\left(\begin{array}{cc}f_i(x)^k & 0\cr
0 &f_i(x)^{k}\end{array}\right)$, \textit{where} $0\leq k\leq 2^\lambda$.

\vskip 2mm \noindent
{\bf 6.} $G_i=\left(\begin{array}{cc}1 & c(x)\cr
0 &f_i(x)^{j}\end{array}\right)$,  \textit{where} $1\leq j\leq 2^\lambda-1$ \textit{and} $c(x)\in A_i/\langle f_i(x)^j\rangle$.

\vskip 2mm \noindent
{\bf 7.} $G_i=\left(\begin{array}{cc}f_i(x)^k & f_i(x)^kc(x)\cr
0 & f_i(x)^{k+j}\end{array}\right)$, \textit{where} $1\leq j\leq 2^\lambda-k-1$,
$1\leq k\leq 2^\lambda-2$ \textit{and} $c(x)\in A_i/\langle f_i(x)^j\rangle$.

\vskip 2mm \noindent
{\bf 8.} $G_i=\left(\begin{array}{cc}c(x) & 1\cr
f_i(x)^{j} & 0\end{array}\right)$, \textit{where} $1\leq j\leq 2^\lambda-1$ \textit{and} $c(x)\in f_i(x)(A_i/\langle f_i(x)^j\rangle)$.

\vskip 2mm \noindent
{\bf 9.} $G_i=\left(\begin{array}{cc}f_i(x)^kc(x) & f_i(x)^k\cr
f_i(x)^{k+j} & 0\end{array}\right)$, \textit{where} $1\leq j\leq 2^\lambda-k-1$,
$1\leq k\leq 2^\lambda-2$ \textit{and} $c(x)\in f_i(x)(A_i/\langle f_i(x)^j\rangle)$.
\end{lemma}

\vskip 3mm
Then from \cite{Norton2000On} Proposition 3.2 and Theorem 3.5, we deduce the following.

\vskip 3mm
   \begin{lemma}\label{la2.1}
\textit{Let $C_i$ be a linear code over $A_i$ of length $2$ with generator matrix
$G_i$}.
\vskip 2mm\par
  (i) \textit{If $G_i$ is given by Cases 1--4 in Lemma \ref{la2.2} and let $\|G_i\|_{f_i(x)}=t$,
then the number of codewords in $C_i$ is equal to $|C_i|=|\mathcal{T}_i|^{2^\lambda-t}=2^{(2^\lambda-t)d_i}$. In this case, we have}
$$C_i=\left\{u(x)G_i\mid u(x)\in \mathbb{F}_2[x]/\langle f_i(x)^{2^\lambda-t}\rangle\right\}.$$

\par
   (ii) \textit{Let $G_i$ be given by Cases 5--9 in Lemma \ref{la2.2} and assume that
$\alpha_1, \alpha_2$ are the two row vectors of $G_i$. If $\|\alpha_k\|_{f_i(x)}=t_k$ for $k=1,2$,
the number of codewords in $C_i$ is equal to $|C_i|=|\mathcal{T}_i|^{(2^\lambda-t_1)+(2^\lambda-t_2))}=2^{(2^{\lambda+1}-t_1-t_2)d_i}$. In this case, we have}
$$C_i=\sum_{k=1,2}\left\{u_k(x)\alpha_k\mid u_k(x)\in \mathbb{F}_2[x]/\langle f_i(x)^{2^\lambda-t_k}\rangle\right\}.$$
\end{lemma}

\par
 $\diamondsuit\diamondsuit$ Then we solve Question 2 at the end of Subsection V.1.
To do this we need the following lemma.

\vskip 3mm
\begin{lemma}\label{la3.1}
\textit{Let $0\leq i\leq r$. Then}

\vskip 2mm\par
  (i) \textit{$x$ is an invertible element of $A_i$ satisfying $x^{-1}=x^{4m-1}$ $({\rm mod} \ f_i(x)^{2^\lambda})$}.

\vskip 2mm\par
  (ii) \textit{In the ring $A_{\mu(i)}$, we have
$$f_i(x^{-1})^k=x^{-k d_i}f_{\mu(i)}(x)^k,$$
where $x^{-kd_i}=x^{4m-k d_i}$ $({\rm mod} \ f_{\mu(i)}(x)^{2^\lambda})$,  for any $1\leq k\leq 2^\lambda-1$. Specifically, we have
$$f_i(x^{-1})^k=x^{-k d_i}f_{i}(x)^k \ {\rm in} \ A_i, \
{\rm when} \ 0\leq i\leq\rho.$$}
\end{lemma}

\begin{IEEEproof}
 (i) Since $f_i(x)^{2^\lambda}$ is a divisor of $x^{4m}-1$ in $\mathbb{F}_2[x]$, we have that
$x^{4m}=1$, i.e. $x^{-1}=x^{4m-1}$ in $A_i$.

\par
  (ii) As $f_i(x)$ is an irreducible divisor of $x^{m_0}-1$ with degree $d_i$, by (i) we see
that $f_i(x^{-1})^k=x^{4m-k d_i}(x^{d_i}f_i(x^{-1}))^k
=x^{4m-k d_i}f_i^\ast(x)^k=x^{4m-k d_i}f_{\mu(i)}(x)^k$ in $A_{\mu(i)}$.
\end{IEEEproof}

\par
  For $A_i$-linear codes $C_i$ of length $2$ satisfying Condition (\ref{eq5}) in Theorem \ref{th2.6},
we have the following theorem.

\vskip 3mm
\begin{theorem}\label{th3.2}
 \textit{Let $0\leq i\leq \rho$. Then all distinct linear codes $C_i$ of length $2$ over $A_i$ satisfying Condition (\ref{eq5}) in Theorem \ref{th2.6} (i) are given by the following
five cases, where $G_i$ is a generator matrix of the code $C_i$}.

\noindent
  I. \textit{$|\mathcal{W}_i^{(2^\lambda)}|$ codes}:
$$G_i=(1, a(x)) \ {\rm with} \ |C_i|=2^{2^\lambda d_i},
\ {\rm where} \ a(x)\in \mathcal{W}_i^{(2^\lambda)}.$$

\noindent
  II. \textit{$\sum_{k=1}^{2^\lambda-1}|\mathcal{W}_i^{(2^\lambda-k)}|$ codes}:
$$G_i=(f_i(x)^k, f_i(x)^k a(x)) \ {\rm with} \ |C_i|=2^{(2^\lambda-k) d_i},$$
\textit{where $a(x)\in \mathcal{W}_i^{(2^\lambda-k)}$ and $1\leq k\leq 2^\lambda-1$}

\noindent
  III. \textit{$2^\lambda+1$ codes}:
$$G_i=f_i(x)^k I_2 \ {\rm with} \ |C_i|=4^{(2^\lambda-k)d_i},$$
  \textit{where $0\leq k\leq 2^\lambda$ and
$I_2$ is the identity matrix of order $2$}.

\noindent
  IV. \textit{$\sum_{j=1}^{2^\lambda-1}|\mathcal{W}_i^{(j)}|$ codes}:
$$G_i=\left(\begin{array}{cc}1 & c(x) \cr 0 & f_i(x)^{j}\end{array}\right)
 \ {\rm with} \ |C_i|=2^{(2^{\lambda+1}-j) d_i},$$
  \textit{where $c(x)\in \mathcal{W}_i^{(j)}$ and $1\leq j\leq 2^\lambda-1$}.

\noindent
  V. \textit{$\sum_{k=1}^{2^\lambda-2}\sum_{j=1}^{2^\lambda-k-1}|\mathcal{W}_i^{(j)}|$ codes}:
$$G_i=\left(\begin{array}{cc}f_i(x)^k & f_i(x)^k c(x) \cr 0 & f_i(x)^{k+j}\end{array}\right)
 \ {\rm with} \ |C_i|=2^{(2^{\lambda+1}-2k-j) d_i},$$
  \textit{where $c(x)\in \mathcal{W}_i^{(j)}$, $1\leq j\leq 2^\lambda-k-1$ and $1\leq k\leq 2^\lambda-2$}.

\vskip 2mm\noindent
  \textit{Therefore, the number of linear codes of length $2$ over $A_i$ satisfying Condition (\ref{eq5})
  in Theorem \ref{th2.6} (i) is equal to}
$$S_{(2,2^{d_i},2^\lambda)}=1+2^\lambda+\sum_{j=1}^{2^\lambda}(2^\lambda-j+1)|\mathcal{W}_i^{(j)}|.$$
\end{theorem}
  \begin{IEEEproof}
See Appendix A.
\end{IEEEproof}

\vskip 3mm\par
   Finally, from Theorems \ref{th2.6} and \ref{th3.2} we deduce the following corollary.

\vskip 3mm
\begin{corollary}\label{co3.3}
\textit{Every binary left $D_{8m}$-code $\mathcal{C}$ can be constructed by the following three steps}:

\vskip2 mm\par
  (i) \textit{For each $i=0,1,\ldots,\rho$, choose a linear code $C_i$ of length $2$ over $A_i$
listed in Theorem \ref{th3.2}}

\vskip2 mm\par
  (ii) \textit{For each $i=\rho+1,\ldots,\rho+\epsilon$, choose a linear code $C_i$ of length $2$ over $A_i$
listed in Lemma \ref{la2.2} and set}
$$C_{i+\epsilon}=\{(b_i(x^{-1}),a_i(x^{-1}))\mid (a_i(x),b_i(x))\in C_i\}\subseteq A_{i+\epsilon}^2.$$

\par
  (iii) \textit{Set
 $$\mathcal{C}=\bigoplus_{i=0}^{r}(\mathcal{A}_i\Box_{\varphi_i}C_i)
 =\sum_{i=0}^r\varepsilon_i(x)C_i \ ({\rm mod} \ x^{4m}-1).$$
 Moreover, the number of codewords in the binary left $D_{8m}$-code $\mathcal{C}$ constructed above is
equal to}
$|{\cal C}|=\prod_{i=0}^{r}|C_i|,$
\textit{where $|C_i|$ is determined by Lemma \ref{la2.1} for all $i=0,1,\ldots,r$}.

\vskip2 mm\noindent
 \textit{Hence the number of all binary left $D_{8m}$-codes is}
$$\left(\prod_{i=0}^\rho S_{(2,2^{d_i},2^\lambda)}\right)\left(\prod_{i=\rho+1}^{\rho+\epsilon} \mathcal{L}_{(2,2^{d_i},2^\lambda)}\right).$$
\end{corollary}

\vskip 3mm\par
  Therefore, in order to list all
distinct binary left $D_{8m}$-codes explicitly, we need to
determine the subset $\mathcal{W}_i^{(s)}$ of
$\mathbb{F}_2[x]/\langle f_i(x)^s\rangle$ precisely, for all
$1\leq s\leq 2^\lambda$ and $0\leq i\leq \rho$. This work has done at
Section III.


\subsection*{V.3 The dual code of every binary left $D_{8m}$-code}
  In this subsection,
we determine the dual code of each binary left $D_{8m}$-code.

\par
    Let $\textit{\textbf{a}}=(a_{0,0}, a_{1,0},\ldots, a_{4m-1,0},a_{0,1}, a_{1,1},\ldots, a_{4m-1,1})$, $\textit{\textbf{b}}=(b_{0,0}, b_{1,0},
   \ldots, b_{4m-1,0},b_{0,1}, b_{1,1},\ldots, b_{4m-1,1})\in \mathbb{F}_2^{8m}$.
The \textit{inner product} of $\textit{\textbf{a}}$ and $\textit{\textbf{b}}$ is defined by
$$[\textit{\textbf{a},\textbf{b}}]=\sum_{i=0}^{4m-1}\sum_{j=0}^1a_{i,j}b_{i,j}\in \mathbb{F}_2.$$
   Let $C$ be a binary linear code of length $8m$, i.e. a subspace of $\mathbb{F}_2^{8m}$. Then the \textit{dual code} of $C$ is defined by
$$C^{\bot}=\{\textit{\textbf{b}}\in \mathbb{F}_2^{8m}\mid [\textit{\textbf{a},\textbf{b}}]=0, \ \forall
\textit{\textbf{a}}\in C\},$$
and $C$ is said to be \textit{self-dual} if
$C=C^{\bot}$.

\par
 For any $a(x)=\sum_{i=0}^{4m-1}a_ix^i\in \mathcal{A}$,
by $x^{4m}=1$ we have
$a(x^{-1})=a_0+\sum_{i=1}^{4m-1}a_ix^{4m-i}\in \mathcal{A}$.
Let $G=\left(g_{ij}(x)\right)_{k\times l}$ be a matrix over $A_i$ of size $k\times l$ with $g_{ij}(x)\in A_i$.
We define
$$\mu(G)=\left(\begin{array}{ccc}g_{11}(x^{-1}) & \ldots & g_{1l}(x^{-1}) \cr \ldots &\ldots & \ldots
\cr g_{k1}(x^{-1}) & \ldots & g_{kl}(x^{-1})\end{array}\right) \
({\rm mod} \ f_{\mu(i)}(x)^{2^\lambda}),$$
and denote  by $G^{{\rm tr}}$ the transpose of $G$, i.e.,
$G^{{\rm tr}}=\left(h_{ij}(x)\right)_{l\times k}$ where $h_{ij}(x)=g_{ji}(x)$.

\vskip 3mm \noindent
\begin{lemma}\label{la4.1}
 \textit{Using the notation above, denote}
$$a_j(x)=\sum_{k=0}^{4m-1}a_{k,j}x^k,\ b_j(x)=\sum_{k=0}^{4m-1}b_{k,j}x^k\in \mathcal{A}, \ j=0,1.$$
\textit{Then $[\textbf{a},\textbf{b}]=0$ if
$(a_0(x),a_1(x))\cdot (b_0(x^{-1}),b_1(x^{-1}))^{\rm tr}
=0$ in the ring $\mathcal{A}=\mathbb{F}_2[x]/\langle x^{4m}-1\rangle$}.
\end{lemma}

\vskip 3mm
   \begin{IEEEproof}
  By $x^{4m}=1$ in $\mathcal{A}$, there exist $h_1,\ldots,h_{4m-1}\in \mathbb{F}_2$ such that
\begin{eqnarray*}
&&(a_0(x),a_1(x))\cdot (b_0(x^{-1}),b_1(x^{-1}))^{\rm tr}\\
 &=&a_0(x)\cdot b_0(x^{-1})+a_1(x)\cdot b_1(x^{-1})\\
 &=&[\textbf{a},\textbf{b}]+h_1x+\ldots+h_{4m-1}x^{4m-1}
 \ ({\rm mod} \ x^{4m}-1).
\end{eqnarray*}
Therefore, $(a_0(x),a_1(x))\cdot (b_0(x^{-1}),b_1(x^{-1}))^{\rm tr}=0$ in $\mathcal{A}$ implies that $[\textbf{a},\textbf{b}]=0$.
\end{IEEEproof}

\vskip 3mm
\par
   Now, we give the dual code of each binary left $D_{8m}$-code.

\vskip 3mm\noindent
\begin{theorem}\label{th4.2}
 \textit{Let $\mathcal{C}$ be a binary left $D_{8m}$-code with
canonical form decomposition ${\cal C}=\bigoplus_{i=1}^{r}(\mathcal{A}_i\Box_{\varphi_i}C_i)$, where
$C_i$ is a linear code of length $2$ over $A_i$ with a generator matrix $G_i$. Then the dual code
of ${\cal C}$ is a binary left $D_{8m}$-code with
the following canonical form decomposition}:
$${\cal C}^{\bot}=\bigoplus_{i=0}^{r}(\mathcal{A}_i\Box_{\varphi_i}Q_i),$$
\textit{where $Q_i$ is a linear code of length $2$ over $A_i$ with a generator matrix $H_i$ given by the following
two cases}:

\vskip 2mm \noindent
  ($\dag$) \textit{Let $0\leq i\leq \rho$. Then $H_i$ is given by one the following
five subcases}:

\vskip 2mm \par
  ($\dag$-I) \textit{$H_i=(1,a(x))$, if $G_i$ is given by Case I in Theorem \ref{th3.2}}.

\vskip 2mm \par
  ($\dag$-II) \textit{$H_i=\left(\begin{array}{cc}1 & a(x) \cr 0 & f_i(x)^{2^\lambda-\kappa}\end{array}\right)$, if $G_i$ is given by Case II in Theorem \ref{th3.2}}.

\vskip 2mm \par
  ($\dag$-III) \textit{$H_i=f_i(x)^{2^\lambda-\kappa} I_2$, if $G_i$ is given by Case III in Theorem \ref{th3.2}}.

\vskip 2mm \par
  ($\dag$-IV) \textit{$H_i=(f_i(x)^{2^\lambda-j},f_i(x)^{2^\lambda-j}c(x))$, if $G_i$ is given by Case IV in Theorem \ref{th3.2}}.

\vskip 2mm \par
  ($\dag$-V) \textit{$H_i=\left(\begin{array}{cc}f_i(x)^{2^\lambda-k-j} & f_i(x)^{2^\lambda-k-j} c(x) \cr  0 & f_i(x)^{2^\lambda-k}\end{array}\right)$,
  if $G_i$ is given by Case V in Theorem \ref{th3.2}}.
\vskip 2mm \noindent
  ($\ddag$) \textit{Let $\rho+1\leq i\leq \rho+\epsilon$. Then the pair $(H_i,H_{i+\epsilon})$
of matrices is given by one the following
nine subcases}:

\vskip 2mm \par
  ($\ddag$-1) \textit{If $G_i$ is given by Case 1 in Lemma \ref{la2.2}},
$$H_i=(1,a(x)) \ {\rm and} \ H_{i+\epsilon}=(a(x^{-1}),1), \
{\rm where} \ a(x)\in A_i.$$

\vskip 2mm \par
  ($\ddag$-2) \textit{If $G_i$ is given by Case 2 in Lemma \ref{la2.2}},
{\small  $$H_i=\left(\begin{array}{cc}1 & a(x) \cr 0 & f_{i}(x)^{2^\lambda-k}\end{array}\right) \ \ {\rm and} \
H_{i+\epsilon}=\left(\begin{array}{cc}a(x^{-1}) & 1 \cr f_{i+\epsilon}(x)^{2^\lambda-k} & 0\end{array}\right),$$ }
\textit{where $a(x)\in A_i/\langle f_i(x)^{2^\lambda-k}\rangle$
and $1\leq k\leq 2^\lambda-1$}.

\vskip 2mm \par
  ($\ddag$-3) \textit{If $G_i$ is given by Case 3 in Lemma \ref{la2.2}},
$$H_i=(f_i(x)b(x),1) \ {\rm and} \ H_{i+\epsilon}=(1,f_{i+\epsilon}(x)\cdot x^{-d_i}b(x^{-1})),$$
\textit{where $b(x)\in A_i/\langle f_i(x)^{2^\lambda-1}\rangle$}.

\vskip 2mm \par
  ($\ddag$-4) \textit{if $G_i$ is given by Case 4 in Lemma \ref{la2.2}},
{\small  $$H_i=\left(\begin{array}{cc} b(x) & 1 \cr f_{i}(x)^{2^\lambda-k} & 0\end{array}\right) \ {\rm and} \
H_{i+\epsilon}=\left(\begin{array}{cc}1 & b(x^{-1}) \cr 0 & f_{i+\epsilon}(x)^{2^\lambda-k} \end{array}\right),$$ }
\textit{where $b(x)\in A_i/\langle f_i(x)^{2^\lambda-k-1}\rangle$
and $1\leq k\leq 2^\lambda-1$}.

\vskip 2mm \par
  ($\ddag$-5) \textit{If $G_i$ is given by Case 5 in Lemma \ref{la2.2}},

\par
 $H_i=\left(\begin{array}{cc} f_i(x)^{2^\lambda-k} & 0\cr 0 & f_{i}(x)^{2^\lambda-k}\end{array}\right)$  \textit{and}

\par
 $H_{i+\epsilon}=\left(\begin{array}{cc}f_{i+\epsilon}(x)^{2^\lambda-k} & 0 \cr 0 & f_{i+\epsilon}(x)^{2^\lambda-k} \end{array}\right),$

\noindent
\textit{where $0\leq k\leq 2^\lambda$}.

\vskip 2mm \par
  ($\ddag$-6) \textit{If $G_i$ is given by Case 6 in Lemma \ref{la2.2}},

\vskip 2mm\par $H_i=(f_i(x)^{2^\lambda-j},f_i(x)^{2^\lambda-j}c(x))$ \textit{and}

\par
  $H_{i+\epsilon}=(f_{i+\epsilon}(x)^{2^\lambda-j}c(x^{-1}),f_{i+\epsilon}(x)^{2^\lambda-j}),$

\vskip 2mm\noindent
 \textit{where $c(x)\in A_i/\langle f_i(x)^{j}\rangle$
and $1\leq j\leq 2^\lambda-1$}.

\vskip 2mm \par
  ($\ddag$-7) \textit{If $G_i$ is given by Case 7 in Lemma \ref{la2.2}},

\par
 $H_i=\left(\begin{array}{cc} f_i(x)^{2^\lambda-k-j} & f_i(x)^{2^\lambda-k-j}c(x)\cr
 0 & f_{i}(x)^{2^\lambda-k}\end{array}\right)$  \textit{and}
 $$H_{i+\epsilon}=\left(\begin{array}{cc}f_{i+\epsilon}(x)^{2^\lambda-k-j}c(x^{-1}) & f_{i+\epsilon}(x)^{2^\lambda-k-j} \cr  f_{i+\epsilon}(x)^{2^\lambda-k} & 0\end{array}\right),$$

\noindent
\textit{where $c(x)\in A_i/\langle f_i(x)^j\rangle$,
$1\leq j\leq 2^\lambda-k-1$ and $1\leq k\leq 2^\lambda-2$}.

\vskip 2mm \par
  ($\ddag$-8) \textit{If $G_i$ is given by Case 8 in Lemma \ref{la2.2}},

\vskip 1mm\par
  $H_i=(f_i(x)^{2^\lambda-j+1}c(x),f_i(x)^{2^\lambda-j})$ \textit{and}

\vskip 1mm\par
 $H_{i+\epsilon}=(f_{i+\epsilon}(x)^{2^\lambda-j},f_{i+\epsilon}(x)^{2^\lambda-j+1}\cdot x^{-d_i}c(x^{-1})),$

\vskip 1mm\noindent
\textit{where $c(x)\in f_i(x)(A_i/\langle f_i(x)^{j}\rangle)$ and $1\leq j\leq 2^\lambda-1$}.

\vskip 2mm \par
  ($\ddag$-9) \textit{If $G_i$ is given by Case 9 in Lemma \ref{la2.2}},

\vskip 1mm\par
 $H_i=\left(\begin{array}{cc} f_i(x)^{2^\lambda-k-j}c(x) & f_i(x)^{2^\lambda-k-j}\cr
f_{i}(x)^{2^\lambda-k} & 0\end{array}\right)$  \textit{and}
 $$H_{i+\epsilon}=\left(\begin{array}{cc}f_{i+\epsilon}(x)^{2^\lambda-k-j}
 & f_{i+\epsilon}(x)^{2^\lambda-k-j}c(x^{-1})
 \cr 0 & f_{i+\epsilon}(x)^{2^\lambda-k} \end{array}\right),$$

\noindent
\textit{where $c(x)\in f_i(x)(A_i/\langle f_i(x)^j\rangle)$,
$1\leq j\leq 2^\lambda-k-1$ and $1\leq k\leq 2^\lambda-2$}.

\end{theorem}

\vskip 3mm\noindent
\begin{IEEEproof}
 For any integer $i$, $0\leq i\leq r$, let $Q_i$ be the linear code of length $2$ over $A_i$ with $H_i$ as its generator matrix. When $0\leq i\leq \rho$, by Theorem \ref{th3.2} we see that $Q_i$ is an $A_i$-submodule of $A_i^2$
satisfying Condition (\ref{eq5}) and $|C_i||Q_i|=2^{2\cdot 2^\lambda d_i}$. When $\rho+1\leq i\leq \rho+2\epsilon$,
by Lemmas \ref{la2.2} and \ref{la2.1} we see that $Q_i$ is an $A_i$-submodule of $A_i^2$
satisfying $|C_i||Q_i|=2^{2\cdot 2^\lambda d_i}$ and
$$Q_{l+\epsilon}=\{(b_l(x^{-1}),a_l(x^{-1}))\mid (a_l(x),b_l(x))\in Q_l\}$$
for all $l=\rho+1,\ldots,\rho+\epsilon$. Now, we set
$\mathcal{Q}=\bigoplus_{i=0}^{r}(\mathcal{A}_i\Box_{\varphi_i}Q_i).$
Then by Theorem \ref{th2.6}, we conclude that
$\mathcal{Q}$ is a binary left $D_{8m}$-code. Moreover, by $4m=2^\lambda\sum_{i=0}^{r}d_i$ and
$|{\cal C}||\mathcal{Q}|=(\prod_{i=0}^{r}|C_i|)(\prod_{i=0}^{r}|Q_i|)=\prod_{i=0}^{r}|C_i||Q_i|$,
we have
\begin{equation}\label{eq7}
|{\cal C}||\mathcal{Q}|=2^{2\cdot 2^\lambda \sum_{i=0}^{r}d_i}
=|\mathbb{F}_2|^{8m}.
\end{equation}

\par
  Let $(a_0(x),a_1(x))\in \mathcal{C}$ and $(b_0(x),b_1(x))\in \mathcal{Q}$. Then there exist
$\xi_i\in A_i$ or $\xi_i\in A^2$ and $\eta_i\in A_i$ or $\eta_i\in A_i^2$ such that

\vskip 1mm\par
 $(a_0(x),a_1(x))=\sum_{i=0}^{r}\varepsilon_i(x)\xi_iG_i\in \mathcal{A}^2,$

\vskip 1mm\par
$(b_0(x),b_1(x))=\sum_{i=0}^{r}\varepsilon_i(x)\eta_iH_i\in \mathcal{A}^2$.

\vskip 1mm\noindent
Then from Lemma \ref{la2.4}(i), Equation (\ref{eq6}) and
$$\mu(i)=\left\{\begin{array}{ll} i, & {\rm when} \ 0\leq i\leq \rho; \cr
                      i+\epsilon, & {\rm when} \ \rho+1\leq i\leq \rho+\epsilon; \cr
                      i-\epsilon, & {\rm when} \ \rho+\epsilon+1\leq i\leq \rho+2\epsilon,\end{array}\right.$$
we deduce that
\begin{eqnarray*}
&&(a_0(x),a_1(x))\cdot (b_0(x^{-1}),b_1(x^{-1}))^{\rm tr}\\
&=&(\sum_{i=0}^{r}\varepsilon_i(x)\xi_iG_i)
 \cdot(\sum_{j=0}^{r}(\varepsilon_j(x^{-1})(\mu(H_j))^{{\rm tr}}(\mu(\eta_j))^{{\rm tr}})\\
&=&(\sum_{i=0}^{r}\varepsilon_i(x)\xi_iG_i)
 \cdot(\sum_{j=0}^{r}(\varepsilon_{\mu(j)}(x)(\mu(H_j))^{{\rm tr}}(\mu(\eta_j))^{{\rm tr}})\\
 &=&\sum_{i=0}^{\rho}\varepsilon_i(x)\xi_i\left(G_i\cdot(\mu(H_i))^{{\rm tr}}\right)(\mu(\eta_i))^{{\rm tr}}\\
 &&+\sum_{i=\rho+1}^{\rho+\epsilon}\varepsilon_i(x)\xi_i\left(G_i\cdot(\mu(H_{i+\epsilon}))^{{\rm tr}}\right)(\mu(\eta_{i+\epsilon}))^{{\rm tr}}\\
 &&+\sum_{i=\rho+\epsilon+1}^{\rho+2\epsilon}\varepsilon_i(x)\xi_i\left(G_i\cdot(\mu(H_{i-\epsilon}))^{{\rm tr}}\right)(\mu(\eta_{i-\epsilon}))^{{\rm tr}}.
\end{eqnarray*}
Now, we have the following two cases.

\par
  {\bf Case ($\dag$)} Let $0\leq i\leq \rho$. By Theorems \ref{th3.2}, we have one of the following subcases:

\par
  Subcase ($\dag$-I) Let $G_i=(1,a(x))$ and $H_i=(1,a(x))$, where $a(x)\in \mathcal{W}_i^{(2^\lambda)}$.
By $a(x)\in \mathcal{W}_i^{(2^\lambda)}$, it follows that $a(x)\in A_i$ satisfying $a(x)a(x^{-1})=1$.
Hence $\mu(H_i)=(1,a(x^{-1}))$ and so
$G_i\cdot(\mu(H_i))^{{\rm tr}}=1+a(x)a(x^{-1})=0.$

\par
  Subcase ($\dag$-II) Let $G_i=(f_i(x)^k,f_i(x)^k a(x))$ and $H_i=\left(\begin{array}{cc}1 & a(x) \cr 0 & f_i(x)^{2^\lambda-k}\end{array}\right)$,
 where $a(x)\in \mathcal{W}_i^{(2^\lambda-k)}$ and $1\leq k\leq 2^\lambda-1$.
Hence
$(\mu(H_i))^{{\rm tr}}=\left(\begin{array}{cc}1 & 0 \cr a(x^{-1}) & f_i(x^{-1})^{2^\lambda-k}\end{array}\right).$
 By $a(x)\in \mathcal{W}_i^{(2^\lambda-k)}$,
it follows that $f_i(x)^k=f_i(x)^k a(x)a(x^{-1})$.
By Lemma \ref{la3.1}(ii), we have
$$f_i(x^{-1})^{2^\lambda-k}=x^{4m-(2^\lambda-k)d_i}f_i(x)^{2^\lambda-k}.$$
Then from a direct calculation, we obtain
$G_i\cdot(\mu(H_i))^{{\rm tr}}=0$.

\par
   One can easily verify that $G_i\cdot(\mu(H_i))^{{\rm tr}}=0$ for Subcases ($\dag$-III), ($\dag$-IV) and ($\dag$-V). Here, we omit the proofs.

\par
  {\bf Case ($\ddag$)} Let $\rho\leq i\leq \rho+2\epsilon$. Then by Theorem \ref{th2.6}, one can easily verify that
$G_i\cdot(\mu(H_{i+\epsilon}))^{{\rm tr}}=0$ for all $i=\rho+1,\ldots,\rho+\epsilon$,
and $G_i\cdot(\mu(H_{i-\epsilon}))^{{\rm tr}}=0$ for all $i=\rho+\epsilon+1,\ldots,\rho+2\epsilon$.

\par
   Therefore, $(a_0(x),a_1(x))\cdot (b_0(x^{-1}),b_1(x^{-1}))^{{\rm tr}}=0$ for any $(a_0(x),a_1(x))\in \mathcal{C}$
and $(b_0(x),b_1(x))\in \mathcal{Q}$. Hence $\mathcal{Q}\subseteq \mathcal{C}^{\bot}$ by Lemma \ref{la4.1}. From this and by Equation (\ref{eq7}), we deduce that
$\mathcal{C}^{\bot}=\mathcal{Q}$ as required.
\end{IEEEproof}

\subsection*{V.4 Proving Theorem \ref{th4.3}}
  We give a proof of Theorem \ref{th4.3} as follows:

\par
   Using the notations of
Theorems \ref{th2.6}, \ref{th3.2} and \ref{th4.2}, let
$${\cal C}=\bigoplus_{i=0}^{r}(\mathcal{A}_i\Box_{\varphi_i}C_i) \ {\rm and}
\ {\cal C}^{\bot}=\bigoplus_{i=0}^{r}(\mathcal{A}_i\Box_{\varphi_i}Q_i),$$
where $C_i$ and $Q_i$ are
$A_i$-submodules of $A_i^2$ determined by Theorem \ref{th4.2}. From this and by Theorem \ref{th2.6},
we deduce that ${\cal C}={\cal C}^{\bot}$ if and only if
for any integer $i$, $0\leq i\leq \rho+\epsilon$, we have that $C_i=Q_i$. The latter is equivalent
to $G_i=H_i$ by Lemma \ref{la2.2}. Hence by
$\lambda\geq 2$ we deduce the following conclusions:

\par
  ($\dag$) Let $0\leq i\leq \rho$. We only need to consider subcase ($\dag$-V) in Theorem \ref{th4.2}. In this case,
$G_i=H_i$ if and only if $k=2^\lambda-k-j$, i.e. $j=2^\lambda-2k$. This implies
$k+j=2^\lambda-k$. Moreover, by $j\geq 1$ we conclude that
$1\leq k\leq 2^{\lambda-1}-1$.

  ($\ddag$) Let $\rho+1\leq i\leq \rho+\epsilon$. Then
the conclusions follows from $G_i=H_i$ and
Theorem \ref{th4.2}($\ddag$) immediately. Moreover, the number of pairs
$(G_i,G_{i+\epsilon})$ is
$$2^{2^\lambda d_i}+2^{(2^\lambda-1) d_i}+1+\sum_{k=1}^{2^\lambda-2}\sum_{j=1}^{2^\lambda-k-1}(2^{jd_i}+2^{(j-1)d_i}),$$
which is equal to $\Omega_{(\lambda,d_i)}$.

\vskip 3mm\noindent
  {\bf Remark} For each self-dual binary left $D_{8m}$-code $\mathcal{C}=\bigoplus_{i=0}^r(\mathcal{A}_i\Box_{\varphi_i}C_i)$
listed by Theorem \ref{th4.3}, a generator matrix of $\mathcal{C}$ is given by
$G_{\mathcal{C}}=\left(\begin{array}{c}B_0\cr B_1\cr\ldots\cr B_r\end{array}\right),$
where $B_i$ is a generator mathix of the subcode $\mathcal{A}_i\Box_{\varphi_i}C_i$, and $B_i$ can be easily determined
by Lemma \ref{la2.1} and the definition of concatenated codes in Section II for all $i=0,1,\ldots,r$.


\section{Conclusion and further work}
Self-dual binary left $D_{8m}$-codes make up an important class of self-dual binary $[8m,4m]$-codes
 such that the dihedral group $D_{8m}$ is necessary a subgroup of the automorphism group of each code.
In this paper, we give an explicit representation and
enumeration for all distinct self-dual binary left $D_{8m}$-codes. In particular,
we provide recursive algorithms to solve
problems in the construction of these codes and obtain a precise formula to count the number of all these codes.
In order to enable readers to use the results of the paper directly to construct self-dual binary $[8m,4m]$-codes,
we give a detailed descriptions of the generator matrices for each self-dual binary left $D_{8m}$-code.

  Future topics of interest include to determine extremal binary self-dual
codes of length $8m$ among self-dual binary left $D_{8m}$-codes, and consider the existence of self-dual, doubly-even and extremal
binary linear codes with basic parameters $[24k, 12k, 4k+4]$ which are also self-dual binary left $D_{24k}$-codes for some integers $k\geq 3$.

\section*{Acknowledgment}
Part of this work was
done when Yonglin Cao was visiting Chern Institute of Mathematics, Nankai
University, Tianjin, China. Yonglin Cao would like to thank the institution
for the kind hospitality. This research is supported in part by the National
Natural Science Foundation of China (Grant Nos. 11671235, 11801324, 61571243, 11701336), the Shandong Provincial Natural Science Foundation, China (Grant No. ZR2018BA007), the Scientific Research Foundation for the PhD of Shandong University of Technology (Grant No. 417037), the Scientific Research Fund of Hubei Provincial Key Laboratory of Applied Mathematics (Hubei University) (Grant No. AM201804) and the Scientific Research Fund of Hunan
Provincial Key Laboratory of Mathematical Modeling and Analysis in
Engineering (No. 2018MMAEZD09).



\section*{Appendix A: Proof of Theorem \ref{th3.2}}

  Let $C_i$ be a linear codes $C_i$ of length $2$ over $A_i$, where $0\leq i\leq \rho$. By Lemma \ref{la2.2}, $C_i$ has one and only one of the following
matrices $G$ as its generator matrix:

\par
   {\bf Cases 1--2} $G=(f_i(x)^k,f_i(x)^k a(x))$, where $0\leq k\leq 2^\lambda-1$ and
   $a(x)\in A_i/\langle f_i(x)^{2^\lambda-k}\rangle$.
   In this case, $C_i$ satisfies Condition (\ref{eq5})
if and only if there exists $u(x)\in A_i$ such that
$$(f_i(x^{-1})^k a(x^{-1}),f_i(x^{-1})^k))=u(x)(f_i(x)^k,f_i(x)^k a(x)),$$
which is equivalent to that
$f_i(x^{-1})^k a(x^{-1})=u(x)f_i(x)^k$ and $f_i(x^{-1})^k=u(x)f_i(x)^k a(x)$. Then we have
$f_i(x^{-1})^k=f_i(x^{-1})^k a(x^{-1}) a(x)$, where $f_i(x^{-1})^k=x^{-kd_i}f_i(x)^k$ by Lemma \ref{la3.1}(ii). From these we deduce that $C_i$ satisfies Condition (\ref{eq5}) if and only if
$a(x)\in A_i/\langle f_i(x)^{2^\lambda-k}\rangle$ satisfying
$$f_i(x)^k (a(x^{-1}) a(x)-1)=0 \ {\rm in} \ A_i.$$
The latter is equivalent
to
$a(x)a(x^{-1})\equiv 1 \ ({\rm mod} \ f_i(x)^{2^\lambda-k})$, i.e., $a(x)\in \mathcal{W}_i^{(2^\lambda-k)}$.

\par
   By $\|(f_i(x)^k,f_i(x)^k a(x))\|_{f_i(x)}=k$ and Lemma \ref{la2.1}(i), it follows that
$|C_i|=2^{(2^\lambda-k) d_i}$.

\par
   {\bf Cases 3--4} $G=(f_i(x)^{k+1}b(x),f_i(x)^k)$, where $0\leq k\leq 2^\lambda-1$ and $b(x)\in A_i/\langle f_i(x)^{2^\lambda-k-1}\rangle$. In this case,
$C_i$ satisfies Condition (\ref{eq5})
if and only if there exists $u(x)\in A_i$ such that
$$(f_i(x^{-1})^k,f_i(x^{-1})^{k+1}b(x^{-1}))=u(x)(f_i(x)^{k+1}b(x),f_i(x)^k).$$
This implies
$f_i(x^{-1})^k=u(x)f_i(x)^{k+1}b(x)$. From this and by $f_i(x^{-1})^k=x^{-kd_i}f_i(x)^k$, we deduce that
$f_i(x)^k=f_i(x)^{k+1}x^{k d_i}u(x)b(x)$. This implies that
$$1\equiv f_i(x)\cdot x^{k d_i}u(x)b(x) \ ({\rm mod} \ f_i(x)^{2^\lambda-k}),$$
and we get a contradiction since $2^\lambda-k\geq 1$. Hence $C_i$ does not satisfy Condition (\ref{eq5}).

\par
   {\bf Case 5} $G=\left(\begin{array}{cc} f_i(x)^k & 0 \cr 0 & f_i(x)^k\end{array}\right)$, where $0\leq k\leq 2^\lambda$.
In this case, we have that $(0, f_i(x^{-1})^k)=x^{4m-k d_i}(0, f_i(x)^k)\in C_i$
and $(f_i(x^{-1})^k,0)=x^{4m-k d_i}(f_i(x)^k,0)\in C_i$ by Lemma \ref{la3.1}(ii). Hence $C_i$ satisfies Condition (\ref{eq5}).

\par
   Moreover, by $\|(f_i(x)^k,0)\|_{f_i(x)}=\|(0,f_i(x)^k)\|_{f_i(x)}=k$ and Lemma \ref{la2.1}(ii), we get
$|C_i|=2^{(2^{\lambda+1}-k-k)d_i}=4^{(2^\lambda-k) d_i}$.

\par
   {\bf Cases 6--7} $G=\left(\begin{array}{cc} f_i(x)^k & f_i(x)^k c(x) \cr 0 & f_i(x)^{k+j}\end{array}\right)$,
where $c(x)\in A_i/\langle f_i(x)^j\rangle$, $1\leq j\leq 2^\lambda-k-1$ and $0\leq k\leq 2^\lambda-2$.
By Lemma \ref{la3.1}(ii), it follows that
\begin{eqnarray*}
(f_i(x^{-1})^{k+j},0)
 &=&(x^{-(k+j)d_i}f_i(x)^{k+j},0)\\
 &=&x^{-(k+j)d_i}f_i(x)^j(f_i(x)^k,f_i(x)^k c(x))\\
 &&-x^{-(k+j)d_i}c(x)(0, f_i(x)^{k+j}).
\end{eqnarray*}
This implies $(f_i(x^{-1})^{k+j},0)\in C_i$.
So $C_i$ satisfies Condition (\ref{eq5}) if and only if there exist $a(x),b(x)\in A_i$ such that
\begin{eqnarray*}
&&(f_i(x^{-1})^k c(x^{-1}),f_i(x^{-1})^k)\\
 &=&a(x)(f_i(x)^k,f_i(x)^k c(x))+b(x)(0, f_i(x)^{k+j}),
\end{eqnarray*}
which is equivalent to $f_i(x^{-1})^k c(x^{-1})=a(x)f_i(x)^k$ and $f_i(x^{-1})^k=a(x)f_i(x)^k c(x)+b(x)f_i(x)^{k+j}$. By
$f_i(x^{-1})^k=x^{-k d_i}$ in $A_i$, the latter condition is equvalent to that
$$f_i(x)^k=f_i(x)^k c(x^{-1})c(x)+x^{k d_i}b(x)f_i(x)^{k+j},$$
i.e., $f_i(x)^k=f_i(x)^k (c(x)c(x^{-1})+x^{k d_i}b(x)f_i(x)^{j})$. This condition is
equivalent to
$$c(x)c(x^{-1})+x^{k d_i}b(x)f_i(x)^{j}\equiv 1 \ ({\rm mod} \ f_i(x)^{2^\lambda-k}).$$
As $1\leq j\leq 2^\lambda-k-1$, we conclude that $C_i$ satisfies Condition (\ref{eq5}) if and only if
$c(x)\in A_i/\langle f_i(x)^j\rangle$ satisfying
$c(x)c(x^{-1})\equiv 1$ (mod $f_i(x)^j$), i.e., $c(x)\in \mathcal{W}_i^{(j)}.$

\par
  Conversely, if $c(x)\in A_i/\langle f_i(x)^j\rangle$ satisfying $c(x)c(x^{-1})\equiv 1$ (mod $f_i(x)^j$), then there exists
element $g(x)\in A_i$ such that $1=c(x)c(x^{-1})+g(x)f_i(x)^j$. Let $b(x)=x^{-k d_i}g(x)\in A_i$. Then we have
$1=c(x)c(x^{-1})+x^{k d_i}b(x)f_i(x)^{j}$. This implies that $f_i(x)^k=f_i(x)^k c(x^{-1})c(x)+x^{k d_i}b(x)f_i(x)^{k+j}$.
From this and by $f_i(x)^{k}=x^{k d_i}f_i(x^{-1})^k$, we deduce that
$f_i(x^{-1})^k=f_i(x^{-1})^k c(x^{-1})c(x)+b(x)f_i(x)^{k+j}$. Select $a(x)=x^{-k d_i}c(x^{-1})\in A_i$. Then
$$a(x)f_i(x)^k=x^{-k d_i}f_i(x)^k c(x^{-1})=f_i(x^{-1})^k c(x^{-1})$$
 and $f_i(x^{-1})^k=a(x)f_i(x)^k c(x)+b(x)f_i(x)^{k+j}$. Hence
\begin{eqnarray*}
&&(f_i(x^{-1})^k c(x^{-1}),f_i(x^{-1})^k)\\
 &=&a(x)(f_i(x)^k,f_i(x)^k c(x))+b(x)(0, f_i(x)^{k+j})\in C_i
\end{eqnarray*}
as required.

\par
   Then by Lemma \ref{la2.1}(ii), $\|(f_i(x)^{k},f_i(x)^{k}c(x))\|_{f_i(x)}=k$ and $\|(0,f_i(x)^{k+j})\|_{f_i(x)}=k+j$,
   it follows that
$|C_i|=2^{(2^{\lambda+1}-(k+j)-k)d_i}=2^{(2^{\lambda+1}-2k-j) d_i}$.

\par
   {\bf Cases 8--9} $G=\left(\begin{array}{cc} f_i(x)^{k}c(x) & f_i(x)^{k} \cr f_i(x)^{k+j} & 0 \end{array}\right)$, where $c(x)\in f_i(x)(A_i/\langle f_i(x)^{j}\rangle)$,
$1\leq j\leq 2^\lambda-k-1$ and $0\leq k\leq 2^\lambda-2$. Then there exists $g(x)\in A_i/\langle f_i(x)^{j}\rangle$ such that $c(x)=f_i(x)g(x)$. This implies $(f_i(x)^{k+1}g(x),f_i(x)^{k})\in C_i$. Suppose that
$(f_i(x^{-1})^{k}, f_i(x^{-1})^{k+1}g(x^{-1}))\in C_i$. Then there exist $a(x), b(x)\in A_i$ such that
\begin{eqnarray*}
&&(f_i(x^{-1})^{k}, f_i(x^{-1})^{k+1}g(x^{-1}))\\
&=&a(x)(f_i(x)^{k+1}g(x), f_i(x)^{k})+b(x)(f_i(x)^{k+j},0).
\end{eqnarray*}
Hence
\begin{eqnarray*}
f_i(x^{-1})^{k}&=&a(x)f_i(x)^{k+1}g(x)+b(x)f_i(x)^{k+j}\\
  &=&f_i(x)^{k+1}\left(a(x)g(x)+b(x)f_i(x)^{j-1}\right).
\end{eqnarray*}
From this and by $f_i(x^{-1})^{k}=x^{-k d_i}f_i(x)^{k}$, we deduce that
$$1\equiv f_i(x)x^{k d_i}\left(a(x)g(x)+b(x)f_i(x)^{j-1}\right)$$
(mod $f_i(x)^{2^\lambda-k}$), where $2^\lambda-k\geq 2$.
But $f_i(x)$ is nilpotent (mod $f_i(x)^{2^\lambda-k}$), we get a contradiction. So $C_i$ does not satisfy Condition (\ref{eq5})
in Theorem \ref{th2.6}.

\par
  Summarizing the above, the number of linear codes  over $A_i$ of length $2$ satisfying Condition (\ref{eq5}) in Theorem \ref{th2.6} (i)
is
$$S_{(2,2^{d_i},2^\lambda)}=1+2^\lambda+\sum_{\kappa=0}^{2^\lambda-1}|\mathcal{W}_i^{(2^\lambda-k)}|
+\sum_{\kappa=0}^{2^\lambda-2}\sum_{j=1}^{2^\lambda-\kappa-1}|\mathcal{W}_i^{(j)}|.$$
After simplifying, we get
$$S_{(2,2^{d_i},2^\lambda)}=1+2^\lambda+\sum_{j=1}^{2^\lambda}(2^\lambda-j+1)|\mathcal{W}_i^{(j)}|.$$

\section*{Appendix B: Results for $\mathcal{W}_1^{(s)}$ ($1\leq s\leq 4$)}

 Let $f_1(x)=x^2+x+1$. For $\mathcal{W}_1^{(s)}$, we have the following calculation results:

{\small \par
 $\mathcal{W}_1^{(1)}=\{1,x,1+x\}$;

\par
 $\mathcal{W}_1^{(2)} =\{1, 1+(1+x)f_1(x), x, x+f_1(x),1+x+f_1(x), 1+x+(1+x)f_1(x)\}$;

\par
 $\mathcal{W}_1^{(3)} =\{1, x^5 + x^3 + x + 1, x^3, x^5 + x, x, x^5 + x^4 + x^3 + x^2 + 1, x^4, x^5 + x^3 + x^2 + x + 1, x^5, x^5 + x^4 + x^2 + 1, x^2, x^4 + 1\}$;

\par
 $\mathcal{W}_1^{(4)}=\{1, x^6 + x^5 + x^3 + x, x^5 + x^3 + x + 1, x^6, x^3, x^6 + x^5 + x + 1, x^5 + x, x^6 + x^3 + 1, x, x^7 + x^6 + x^4 + x^2, x^7, x^6 + x^4 + x^2 + x, x^4, x^7 + x^6 + x^2 + x, x^7 + x^4 + x, x^6 + x^2, x^5, x^7 + x^4 + x^3 + x^2 + 1, x^5 + x^4 + x^2 + 1, x^7 + x^3, x^2, x^7 + x^5 + x^4 + x^3 + 1, x^4 + 1, x^7 + x^5 + x^3 + x^2\}$. }

\section*{Appendix C: Expressions for $a(x)$ and $b(x)$ in Example \ref{ex6.3}}

\vskip 2mm \noindent
   {\bf Case 1} $a(x)=x^4+x^2+1$ and $b(x)=\sum_{j=0}^{15}b_jx^j$, where
$b_0b_1b_2\ldots b_{15}\in \mathbb{F}_2^{16}$ is given by one of the following $24$ cases:

\vskip 1mm
{\small
$1001011100010100$, $1000001101010001$, $1001001101010000$,

$1000011100010101$, $0100001000001101$, $1001011001001000$,

$0100011001001001$, $0101001000001100$, $0001111001010110$,

$0100111101000010$, $0000111101000110$, $0101111001010010$,

$1000001110010111$, $1101001010000011$, $1001001010000111$,

$1100001110010011$, $0000110001011110$, $0100100100001111$,

$0100100000011110$, $0000110101001111$, $0001010010000011$,

$0101000111010010$, $0101000011000011$, $0001010110010010$.   }

\vskip 2mm \noindent
   {\bf Case 2} $a(x)=x^6 + x^4 + x^2$ and $b(x)=\sum_{j=0}^{15}b_jx^j$, where
$b_0b_1b_2\ldots b_{15}\in \mathbb{F}_2^{16}$ is given by one of the following $24$ cases:

\vskip 1mm
{\small
$1001011100010100$, $1000001101010001$, $1001001101010000$,

$1000011100010101$, $1111010001100000$, $1110000000100101$,

$1111000000100100$, $1110010001100001$, $1100010111100000$,

$1001010011110100$, $1101010011110000$, $1000010111100100$,

$1000001110010111$, $1101001010000011$, $1001001010000111$,

$1100001110010011$, $0110000110000101$, $0010010011010100$,

$0010010111000101$, $0110000010010100$, $0001010010000011$,

$0101000111010010$, $0101000011000011$, $0001010110010010$.  }

\vskip 2mm \noindent
   {\bf Case 3} $a(x)=x^6 + x^4 + 1$ and $b(x)=\sum_{j=0}^{15}b_jx^j$, where
$b_0b_1b_2\ldots b_{15}\in \mathbb{F}_2^{16}$ is given by one of the following $24$ cases:

\vskip 1mm
{\small
$0010000101111001$, $0011010100111100$, $0010010100111101$,

$0011000101111000$, $1111010001100000$, $1110000000100101$,

$1111000000100100$, $1110010001100001$, $1100010111100000$,

$1001010011110100$, $1101010011110000$, $1000010111100100$,

$0101100000100001$, $0000100100110101$, $0100100100110001$,

$0001100000100101$, $0110000110000101$, $0010010011010100$,

$0010010111000101$, $0110000010010100$, $0111100101011000$,

$0011110000001001$, $0011110100011000$, $0111100001001001$. }

\vskip 2mm \noindent
   {\bf Case 4} $a(x)=x^6 + x^2 + 1$  and $b(x)=\sum_{j=0}^{15}b_jx^j$, where
$b_0b_1b_2\ldots b_{15}\in \mathbb{F}_2^{16}$ is given by one of the following $24$ cases:

\vskip 1mm
{\small
$0010000101111001$, $0011010100111100$, $0010010100111101$,

$0011000101111000$, $0100001000001101$, $0101011001001000$,

$0100011001001001$, $0101001000001100$, $0001111001010110$,

$0100111101000010$, $0000111010001100$, $0101111001010010$,

$0101100000100001$, $0000100100110101$, $0100100100110001$,

$0001100000100101$, $0000110001011110$, $0100100100001111$,

$0100100000011110$, $0000110101001111$, $0111100101011000$,

$0011110000001001$, $0011110100011000$, $0111100001001001$. }

\vskip 2mm \noindent
   {\bf Case 5} $a(x)=x^5 + x^3 + x$ and $b(x)=\sum_{j=0}^{15}b_jx^j$, where
$b_0b_1b_2\ldots b_{15}\in \mathbb{F}_2^{16}$ is given by one of the following $24$ cases:

\vskip 1mm
{\small
$0010100011101001$, $1000101011000001$, $0000101011001001$,

$1010100011100001$, $0000011000101111$, $1010010000000111$,

$0010010000001111$, $1000011000100111$, $0100101110001010$,

$1100000100101000$, $1100001100001010$, $0100100110101000$,

$1010000110000110$, $0010101100100100$, $0010100100000110$,

$1010001110100100$, $0000011110100011$, $0010111100101001$,

$0000111100101011$, $0010011110100001$, $1100000101001011$,

$1110100111000001$, $1100100111000011$, $1110000101001001$. }

\vskip 2mm \noindent
   {\bf Case 6} $a(x)=x^7 + x^5 + x^3$ and $b(x)=\sum_{j=0}^{15}b_jx^j$, where
$b_0b_1b_2\ldots b_{15}\in \mathbb{F}_2^{16}$ is given by one of the following $24$ cases:

\vskip 1mm
{\small
$0010100011101001$, $1000101011000001$, $0000101011001001$,

$1010100011100001$, $1011000001000010$, $0001001001101010$,

$1001001001100010$, $0011000001001010$, $0100101110001010$,

$1100000100101000$, $1100001100001010$, $0100100110101000$,

$0111101000110000$, $1111000010010010$, $1111001010110000$,

$0111100000010010$, $0110101001111000$, $0100001011110010$,

$0110001011110000$, $0100101001111010$, $1100000101001011$,

$1110100111000001$, $1100100111000011$, $1110000101001001$. }

\vskip 2mm \noindent
   {\bf Case 7} $a(x)=x^7 + x^5 + x$ and $b(x)=\sum_{j=0}^{15}b_jx^j$, where
$b_0b_1b_2\ldots b_{15}\in \mathbb{F}_2^{16}$ is given by one of the following $24$ cases:

\vskip 1mm
{\small
$1001111010000100$, $0011110010101100$, $1011110010100100$,

$0001111010001100$, $1011000001000010$, $0001001001101010$,

$1001001001100010$, $0011000001001010$, $1001000000111100$,

$0001101010011110$, $0001100010111100$, $1001001000011110$,

$0111101000110000$, $1111000010010010$, $1111001010110000$,

$0111100000010010$, $0110101001111000$, $0100001011110010$,

$0110001011110000$, $0100101001111010$, $1010110010010000$,

$1000010000011010$, $1010010000011000$, $1000110010010010$. }

\vskip 2mm \noindent
   {\bf Case 8} $a(x)=x^7 + x^3 + x$ and $b(x)=\sum_{j=0}^{15}b_jx^j$, where
$b_0b_1b_2\ldots b_{15}\in \mathbb{F}_2^{16}$ is given by one of the following $24$ cases:

\vskip 1mm
{\small
$1001111010000100$, $0011110010101100$, $1011110010100100$,

$0001111010001100$, $0000011000101111$, $1010010000000111$,

$0010010000001111$, $1000011000100111$, $1001000000111100$,

$0001101010011110$, $0001100010111100$, $1001001000011110$,

$1010000110000110$, $0010101100100100$, $0010100100000110$,

$1010001110100100$, $0000011110100011$, $0010111100101001$,

$0000111100101011$, $0010011110100001$, $1010110010010000$,

$1000010000011010$, $1010010000011000$, $1000110010010010$. }

\section*{Appendix D: Expressions for $a(x)$ and $\eta(x)$ in Example \ref{ex6.4}}

\vskip 2mm \noindent
   {\bf Case 1} $a(x)\in\{x^2 + x + 1, x^3 + x^2 + 1, x^3 + x^2 + x, x^3 + x + 1\}$
 and $\eta(x)=\sum_{j=0}^{11}\eta_jx^j$, where
$\eta_0\eta_1\eta_2\ldots \eta_{11}\in \mathbb{F}_2^{12}$ is given by one of the following $168$ cases:

\vskip 1mm
{\small
$000000001101$, $000000010111$, $000000011010$, $000000101110$,

$000000110100$, $000000111001$, $000001011100$, $000001100101$,

$000001101000$, $000001110010$, $000010011011$, $000010111000$,

$000011001010$, $000011010000$, $000011011101$, $000011100100$,

$000100110110$, $000100111011$, $000101001001$, $000101010011$,

$000101100111$, $000101110000$, $000110000011$, $000110010100$,

$000110100000$, $000110111010$, $000111000101$, $000111001000$,

$001000001001$, $001000100111$, $001001010101$, $001001100001$,

$001001101100$, $001001110110$, $001010010010$, $001010100110$,

$001010101011$, $001010110001$, $001011001110$, $001011100000$,

$001100000110$, $001100001011$, $001100100101$, $001100101000$,

$001101000000$, $001101010111$, $001101100011$, $001101110100$,

$001110001010$, $001110010000$, $001110101001$, $001110110011$,

$010000010010$, $010000110001$, $010001000011$, $010001001110$,

$010001011001$, $010001110111$, $010010001001$, $010010101010$,

$010011000010$, $010011010101$, $010011011000$, $010011101100$,

$010100000111$, $010100011101$, $010100100100$, $010100110011$,

$010101001100$, $010101010110$, $010101100010$, $010101110101$,

$010110010001$, $010110011100$, $010111000000$, $010111001101$,

$011000000001$, $011000001100$, $011000010110$, $011000110101$,

$011001000111$, $011001001010$, $011001010000$, $011001110011$,

$011010000000$, $011010101110$, $011011000110$, $011011101000$,

$011100000011$, $011100010100$, $011100100000$, $011100110111$,

$011101000101$, $011101010010$, $011101100110$, $011101110001$,

$100000011101$, $100000100100$, $100000101001$, $100000110011$,

$100001000001$, $100001100010$, $100010000110$, $100010001011$,

$100010011100$, $100010110010$, $100011001101$, $100011101110$,

$100100000101$, $100100010010$, $100100101011$, $100100110001$,

$100101010100$, $100101011001$, $100110000100$, $100110010011$,

$100110101010$, $100110110000$, $100111010101$, $100111011000$,

$101000000011$, $101000001110$, $101000011001$, $101000111010$,

$101001001000$, $101001100110$, $101010011000$, $101010100001$,

$101010101100$, $101010111011$, $101011000100$, $101011101010$,

$101100000001$, $101100011011$, $101100100010$, $101100111000$,

$101110000000$, $101110011010$, $101110100011$, $101110111001$,

$110000000010$, $110000010101$, $110000011000$, $110000101100$,

$110001001001$, $110001101010$, $110010001110$, $110010010100$,

$110010011001$, $110010100000$, $110011000101$, $110011100110$,

$110100000000$, $110100001101$, $110101010001$, $110101011100$,

$110110000001$, $110110001100$, $110111010000$, $110111011101$,

$111000000110$, $111000101000$, $111001000000$, $111001101110$,

$111010001010$, $111010100100$, $111011001100$, $111011100010$.  }

\vskip 2mm \noindent
   {\bf Case 2} $a(x)=x^2 + x + 1$ and $\eta(x)=\sum_{j=0}^{11}\eta_jx^j$, where
$\eta_0\eta_1\eta_2\ldots \eta_{11}\in \mathbb{F}_2^{12}$ is given by one of the following $14$ cases:

\vskip 1mm
{\small
$000001000110$, $000010001100$, $001000000100$, $001001000010$,

$010000001000$, $010000100110$, $010001100000$, $010010000100$,

$011000100010$, $011001100100$, $100001001100$, $100011000000$,

$110001000100$, $110011001000$.  }

\vskip 2mm \noindent
   {\bf Case 3} $a(x)=x^3 + x^2 + 1$ and $\eta(x)=\sum_{j=0}^{11}\eta_jx^j$, where
$\eta_0\eta_1\eta_2\ldots \eta_{11}\in \mathbb{F}_2^{12}$ is given by one of the following $14$ cases:

\vskip 1mm
{\small
$000000100011$, $000010000001$, $000100000010$, $000100011000$,

$000100100001$, $000110011001$, $001000010011$, $001000110000$,

$001100010001$, $001100110010$, $100000010000$, $100010010001$,

$100100001000$, $100110001001$. }

\vskip 2mm \noindent
   {\bf Case 4} $a(x)=x^3 + x^2 + x$ and $\eta(x)=\sum_{j=0}^{11}\eta_jx^j$, where
$\eta_0\eta_1\eta_2\ldots \eta_{11}\in \mathbb{F}_2^{12}$ is given by one of the following $14$ cases:

\vskip 1mm
{\small
$000000100011$, $000001000110$, $000100000010$, $000100100001$,

$001000000100$, $001000010011$, $001000110000$, $001001000010$,

$001100010001$, $001100110010$, $010000100110$, $010001100000$,

$011000100010$, $011001100100$. }

\vskip 2mm \noindent
   {\bf Case 5} $a(x)=x^3 + x + 1$ and $\eta(x)=\sum_{j=0}^{11}\eta_jx^j$, where
$\eta_0\eta_1\eta_2\ldots \eta_{11}\in \mathbb{F}_2^{12}$ is given by one of the following $14$ cases:

\vskip 1mm
{\small
$000010000001$, $000010001100$, $000100011000$, $000110011001$,

$010000001000$, $010010000100$, $100000010000$, $100001001100$,

$100010010001$, $100011000000$, $100100001000$, $100110001001$,

$110001000100$, $110011001000$.  }

\ifCLASSOPTIONcaptionsoff
  \newpage
\fi



%
\bibliographystyle{IEEEtran}
\bibliography{bibfile}

%






\end{document}